\documentclass[preprint,aps,nofootinbib,prd,showpacs]{revtex4}
\usepackage[colorlinks=true,linkcolor=blue,urlcolor=blue,filecolor=black,citecolor=red,pdfstartview=FitV,pdftitle={},pdfsubject={},pdfkeywords={},pdfpagemode=None,bookmarksopen=true]{hyperref}
\usepackage{graphicx}
\usepackage{amsmath}
\usepackage{amsfonts}
\usepackage{amssymb,ulem}
\usepackage{color}%
\usepackage{slashed}

\providecommand{\U}[1]{\protect\rule{.1in}{.1in}}

\begin{document}
\title{Anomaly freedom in perturbative models\\ of Euclidean loop quantum
  gravity} \author{Jian-Pin Wu $^{1}$}
\email{jianpinwu@mail.bnu.edu.cn} \author{Martin Bojowald $^{2}$}
\email{bojowald@gravity.psu.edu} \author{Yongge Ma $^{3}$}
\email{mayg@bnu.edu.cn} \affiliation{$^1$Center for Gravitation and Cosmology, College of Physical Science and Technology,
Yangzhou University, Yangzhou 225009, China\\
  $^2$ Institute for Gravitation and the Cosmos, The Pennsylvania State
  University, University Park, PA 16802, USA\\
  $^3$ Department of Physics, Beijing Normal University, Beijing, 100875,
  China}

\begin{abstract}
  Euclidean gravity provides an interesting test system for an analysis of
  cosmological perturbations in an effective Hamiltonian constraint with
  holonomy modifications from loop quantum gravity. This paper presents a
  discussion of scalar modes, with a specific form of the holonomy
  modification function derived from a general expansion in a connection
  formulation.  Compared with some previous models, the constraint brackets
  are deformed in a different and more restricted way. A general comparison of
  anomaly-free brackets in various effective and operator versions shows
  overall consistency between different approaches.
\end{abstract}


\maketitle

\tableofcontents

\section{Introduction}

Loop quantum gravity \cite{Rov,ThomasRev,ALRev,LQGHan}, implements
non-perturbative and background-independent features in an approach to
quantizing general relativity. It could therefore provide models of
quantum space-time structure. To this end, one should address the
long-standing anomaly problem of space-time gauge transformations in order to
shed light on consistent versions. Without such a derivation, assuming certain
properties of solutions, for instance in the form of an effective line
element, amounts to postulating a background space-time.  Although overall
consistency of the theory remains to be shown, there are now several
encouraging results which indicate that a well-defined quantum space-time
structure may be realized. If this is the case, one could potentially use the
theory to derive possible effects for instance in cosmological observations.

In addition to a consistent theory, a systematic effective framework is
required for a reliable evaluation of physical phenomena. In the
background-independent context of loop quantum gravity, such methods have been
explored by both the canonical
\cite{BouncePert,EffRecollapse,AltEffRecollapse,EffConsRel,EffConsComp,HigherMoments}
and the path integral perspective
\cite{PathIntEff,PathIntAlt,FuncCollapse1,FuncCollapse2} in homogeneous
models. For inhomogeneous modes of cosmological perturbations, one encounters
new questions related to the consistency of coupled partial differential
equations, or the anomaly-problem of quantum gravity.

In order to understand cosmological structure formation and anisotropies of
the cosmic microwave background in models of loop quantum gravity, one
needs to consider a cosmological perturbation theory with modifications
including quantum-gravity effects.  In the canonical setting of loop quantum
gravity, quantum-gravity effects appear in an effective Hamiltonian
constraint, rather than an effective action whose covariance could be checked
directly. If the corrections implied by a canonical theory of quantum gravity
are not covariant, Hamiltonian (and diffeomorphism) constraints obey
constraint brackets which do not close but rather contain anomaly terms
$\mathcal{A}_{IJ}$: Poisson brackets of two constraints would not weakly
vanish but be of the form
\begin{equation} \label{CAQuantumLevel}
\{\mathcal{C}_{I},\mathcal{C}_{J}\}=\mathcal{K}^{K}_{IJ}
\mathcal{C}_{K} + \mathcal{A}_{IJ}
\end{equation}
with $\mathcal{A}_{IJ}\not=0$.  If there is such an anomaly, the quantum
corrected perturbation equations cannot be expressed solely in terms of
gauge-invariant variables \cite{ConstraintAlgebra}.  Therefore, how to obtain
anomaly-free constraints of cosmological perturbations including
loop quantum effects has become an important question.

Several promising results have been obtained in this direction, exploring the
commutators of constraint operators
\cite{ThreeDeform,TwoPlusOneDef,TwoPlusOneDef2,AnoFreeWeak,SphSymmOp,OffShell,ConstraintsG}
or Poisson brackets of effective constraints
\cite{ConstraintAlgebra,JR,ScalarHol,ScalarHolInv,LTBII,HigherSpatial,SphSymmComplex,CosmoComplex,GowdyComplex}. In
models analyzed so far, it seems possible to have closed brackets
($\mathcal{A}_{IJ}=0$), but usually with modifications of the structure
functions $\mathcal{K}^K_{IJ}$ in (\ref{CAQuantumLevel}), in particular for
real connections. The classical brackets corresponds to a canonical version of
space-time coordinate transformations, represented as deformations of spatial
hypersurfaces in space-time \cite{Regained}. If the brackets are modified (and
not just its generators), the gauge transformations generated by the
constraints are not broken but differ from coordinate transformations, so that
a new space-time model is obtained. Only in some cases may it be possible to
map the effective geometry to one of classical type by applying a field
redefinition \cite{Normal,EffLine}. The most dramatic effect found in this
context is the possibility of signature change
\cite{Action,SigChange,SigImpl,PhysicsToday} at large density or curvature,
indicated by a change of sign in some of the structure functions. Such an
effect is interesting, but also dangerous owing to the indeterministic
behavior that it may imply \cite{Loss}. In this article, we consider a model
which turns out to lead to different implications in situations that would
give rise to signature change in previous models. In this respect, our results
are related to those of \cite{SphSymmComplex,CosmoComplex,GowdyComplex}, but
qualitatively they are obtained in a different way.

In general, there are two main quantum-gravity effects in loop-quantized
models, so-called inverse-triad corrections \cite{InvScale,Ambig} and holonomy
modifications \cite{EffHam,AmbigConstr}. In addition to these two, there are
generic quantum back-reaction effects which occur in all interacting quantum
theories but have not been explored much in inhomogeneous models of loop
quantum gravity. We will continue this tradition and mostly ignore these terms
in the present paper, focussing on the two types of corrections directly
related to quantum geometry. (As shown in \cite{EffConsQBR}, under certain
conditions quantum back-reaction terms from moments do not appear in structure
functions of constraint brackets.) For the case of inverse-triad corrections,
anomaly-free constraints and the corresponding gauge-invariant cosmological
perturbation equations have been obtained for scalar modes
\cite{ConstraintAlgebra,ScalarGaugeInv}, vector modes \cite{Vector} and tensor
modes \cite{Tensor}, respectively. (For tensor modes, anomaly-freedom of the
constraints is automatically fulfilled.) A characteristic feature, shared with
spherically symmetric models, is that the Poisson bracket of two Hamiltonian
constraints is modified by a factor of the square of the inverse-triad
correction function. As this function is positive, signature change does not
happen. Some relevant applications, including potentially observable effects
in the primordial power spectrum and non-Gaussianity, have already been
studied \cite{LoopMuk,InflConsist,InflTest,NonGaussInvVol}.

Holonomy modifications have been implemented in consistent versions slightly
more recently. The first papers used a partial gauge fixing to longitudinal
gauge \cite{CosPertHolLong,ScalarHolEv} and therefore could not show all
effects with full confidence. Without gauge fixing, a consistent version has
been obtained in \cite{VectorHol} for vector modes and \cite{ScalarHol} for
scalar modes.  A combined treatment of holonomy-modified scalar, vector and
tensor perturbations has been given in \cite{ScalarTensorHol}. Again,
anomaly-free constraints can be obtained by a rather simple quantum correction
for all types of perturbations. In the presence of holonomy modifications, the
constraint brackets are modified in such a way that structure functions may
change sign, corresponding to a transition between Lorentzian and Euclidean
signature in the sense that either hyperbolic or elliptic mode equations are
implied \cite{Action,SigChange,SigImpl}. There is agreement with consistent
constraint brackets in spherically symmetric models
\cite{JR,LTBII,HigherSpatial} even at the operator level
\cite{SphSymmOp}. (See \cite{DeformedCosmo} for a comparison.)  Signature
change is not always realized in self-dual variables
\cite{SphSymmComplex,CosmoComplex,GowdyComplex} because the Hamiltonian
constraint has a different formal structure in its dependence on spatial
derivatives of the fields.

Anomaly-free constraints for both inverse-triad and holonomy modifications
have been studied for all types of perturbative modes. The corresponding
equations of motion are derived in \cite{ScalarHolInv}, providing so far the
most complete treatment of consistent cosmological perturbations in models of
loop quantum cosmology.  However, in a certain sense, holonomy modifications
so far have been considered after rather than before perturbing the classical
Hamiltonian constraint: One modifies the background constraint by replacing
the classical quadratic dependence on the connection $\bar{q}$ (or Hubble
parameter) by a bounded function, $\bar{q}^2\mapsto
\ell^{-2}\sin^2(\ell\bar{q})$, as it has been found by effective equations of
isotropic models \cite{EffHam}, and then looks for a possible anomaly-free
theory of perturbative modes on such a background model.  If one perturbs a
modified constraint, additional terms may appear. In particular, there could
be derivative corrections, even at or below the classical derivative order,
which happen to be absent in the classical constraints but might be induced by
quantum-geometry effects. (See \cite{HigherSpatial} for a discussion in
spherical symmetry.) In covariant effective actions, all quantum corrections
are expected to be of higher-derivative (or higher-curvature) type, but
lower-order terms may appear if the space-time structure is modified as in
certain canonical approaches. An effective treatment should include all terms,
up to a given order, consistent with what is known about symmetries. If the
precise form of quantum space-time is unknown, one cannot assume much about
symmetries and should include all possible terms in an ansatz for an effective
Hamiltonian. Symmetries will then be implemented by the condition of anomaly
freedom, and their possible form can be derived from the effective system
rather than being assumed. By including additional derivative terms, we
therefore fill in a gap in existing treatments.

In a canonical setting, the treatment of spatial and temporal derivatives is
different. The former appear directly in an effective Hamiltonian while the
latter would result in an adiabatic approximation of quantum back-reaction
\cite{EffAc,Karpacz,HigherTime}. Although both types of derivatives should
usually be considered in combination, holonomy modifications suggest a larger
role for spatial derivatives because holonomies are spatially non-local
functions of the connection. If holonomy modifications can be consistent in
cosmological perturbation theory, one should therefore be able to find
anomaly-free constraints with holonomy modifications of the background and a
set of higher spatial derivative terms.

In order to explore the perturbations in a framework including holonomy
modifications of loop quantum gravity, allowing for more general derivative
terms than considered in \cite{ScalarHolInv}, an effective holonomy-modified
Hamiltonian in Euclidean general relativity was first proposed in
\cite{VectorSecond}, where the corresponding perturbative constraint brackets
were studied for vector modes.  The Poisson brackets between the modified
Hamiltonian and diffeomorphism constraints restricted to vector modes were
calculated, and a specific form of the holonomy-modification function
$f^{i}_{cd}$ giving rise to anomaly-free constraints was found.  This result
indicates that in a perturbative framework it is indeed possible to have
non-trivial and anomaly-free holonomy modifications with additional derivative
terms up to first order, as suggested by non-local holonomies in the full
theory.  In this paper, we shall extend the study to scalar modes in the same
framework.

A brief review of the modification function of the full theory and some basic
elements of scalar modes will be presented in section \ref{SReview}.  Then, in
section \ref{ConstraintAlgebra}, the constraint brackets, including those
between the modified Hamiltonian constraint and the diffeomorphism constraint
as well as between the two modified Hamiltonian constraints, are derived.
Subsequently, a specific form of the holonomy modification function is
obtained from its general expression in section \ref{Form1}.  We compare the
results with those of \cite{ScalarHolInv} on one hand, and those of
\cite{SphSymmComplex,CosmoComplex,GowdyComplex} on the other, and discuss
implications for signature change in section \ref{Conclusions}. Results from
operator approaches are briefly discussed as well.

At a formal level, the difference between \cite{ScalarHolInv} and our present
treatment is that we use a connection formulation and include additional
derivative terms of the connection. Interestingly, the outcome does not seem
to be the same. Our calculations lead to an intermediate set of deformed
constraint brackets which may show a way to avoid signature change and the
associated indeterministic behavior, but we have not been able to produce a
fully consistent non-classical system: While the brackets of Hamiltonian and
diffeomorphism constraints can be closed, the expressions are not
SU(2)-covariant unless there are no holonomy corrections (while inverse-triad
corrections may be possible). We interpret this result as an indication that
non-local modifications are essential in SU(2)-invariant connection theories.

\section{Holonomy modification functions and scalar modes}\label{SReview}

In the connection formulation of Euclidean general relativity
\cite{ALRev,LQGHan}, the gravitational Hamiltonian constraint
can be written as
\begin{eqnarray}
\label{GravityHamiltonian}
H[N]=\frac{1}{16\pi G}\int _{\Sigma}{\rm d}^{3}x N
\epsilon^{\ jk}{}_{i}\frac{E^{c}_{j}E^{d}_{k}}{\sqrt{|\det E|}}F^{i}_{cd}\,,
\end{eqnarray}
where $E^b_j$ is the densitized triad, and the curvature of the
Ashtekar--Barbero connection $A^i_a=\Gamma_a^i+K_a^i$ is given by
\begin{eqnarray} \label{Curvature}
F^{i}_{cd}=2\partial_{[c}A^{i}_{d]}
+\epsilon^{i}_{\ mn}A^{m}_{c}A^{n}_{d}\,.
\end{eqnarray}
In the expression for $A_a^i$, $\Gamma_a^i$ is the spin connection compatible
with the triad, and $K_a^i=K_{ab}E^{bi}/|\det E^c_j|$ is obtained from
extrinsic curvature $K_{ab}$. More generally, one can define
$A_a^i=\Gamma_a^i+\gamma K_a^i$ with the Barbero--Immirzi parameter $\gamma$
\cite{AshVarReell,Immirzi}. If $\gamma\not=1$, there will be additional terms
in the Hamiltonian constraint which contain spatial derivatives of the
densitized triad, on whose relevance we will comment later. We use the value
$\gamma=1$ in order to work with the simplified expression
(\ref{GravityHamiltonian}).

In loop quantum gravity, the local dependence on the connection $A^{i}_{a}$ is
replaced by a dependence on non-local (in space) holonomies
\begin{eqnarray}
\label{Holonomy}
h_{e}(A)=\mathcal{P}\exp\int_{e}A^{i}_{a}\tau_{i}{\rm d}x^{a}
\end{eqnarray}
for suitable choices of spatial curves $e$, where the symbol $\mathcal{P}$
represents path ordering, and $\tau_{j}=-\frac{i}{2}\sigma_{j}$ is a basis of
the Lie-algebra su(2) with $\sigma_{j}$ being the Pauli matrices.  Holonomies,
unlike connection components, can be represented as operators on the
kinematical Hilbert space of loop quantum gravity, and therefore appear in
candidates for the quantized Hamiltonian constraint \cite{RS:Ham,QSDI}.

However, it is difficult to find anomaly-free versions because the operators
and their commutators are complicated expressions depending sensitively on
factor orderings and other quantization choices. There has been some progress
in particular but not only in $2+1$-dimensional models
\cite{ThreeDeform,TwoPlusOneDef,TwoPlusOneDef2,AnoFreeWeak,OffShell,ConstraintsG},
with consistent commutators on a subset of states which partially solve the
spatial diffeomorphism constraint (introduced in \cite{LM:Vertsm,Consist}).
Attempts to go beyond the restricted set of states
\cite{OffShell,ConstraintsG} in Euclidean gravity indicate that closed
commutators of constraint operators may be possible more generally.
Unfortunately, the complicated semiclassical limit of such theories makes it
difficult to see the full implications of holonomy modifications, in
particular those related to potential deformations of the constraint brackets
and signature change.

An effective approach to constraints has proven to be more powerful
\cite{EffCons,EffConsRel,EffConsQBR}, in which one does not directly compute
commutators $[\hat{C}_I,\hat{C}_J]$ of constraint operators but rather Poisson
brackets
\begin{equation}
 \{\langle\hat{C}_I\rangle,\langle\hat{C}_J\rangle\}:=
\frac{\langle[\hat{C}_I,\hat{C}_J]\rangle}{i\hbar}
\end{equation}
of effective constraints $\langle\hat{C}_I\rangle$. Methods have been
developed by which one can evaluate the left-hand side in an expansion by
quantum moments, which turns out to be more feasible than computing quantum
commutators. These methods, applied to a fixed order in $\hbar$, cannot show
whether a consistent operator version exists. But they can rule out certain
choices, or provide indications of necessary deformations of the brackets when
certain modifications, such as holonomy terms, are to be implemented. (For a
general discussion, see \cite{EffConsQBR}.)  So far, the expansions used in
the context of cosmological perturbations have been done to lowest order in
$\hbar$, which means that one ignores quantum back-reaction but allows for
some quantum-geometry effects.

In order to include holonomy modifications in an effective theory of this form,
we could, in general, consider the following ansatz of holonomy modifications to
the Euclidean Hamiltonian \cite{VectorSecond}
\begin{eqnarray}
\label{EuclideanHdwithHolonomy}
\delta \mathcal{H}_Q=\epsilon^{jk}{}_{i}\frac{E^{c}_{j}E^{d}_{k}}{\sqrt{|\det
    E|}}f^{i}_{cd}
(A,\partial A,\partial^{2} A,\cdots,\partial^{n} A,\epsilon)\,,
\end{eqnarray}
where $f^{i}_{cd}(A,\partial A,\partial^{2} A,\cdots,\partial^{n}
A,\epsilon)+O(\partial^{n+1}A)= \tilde{F}^{i}_{cd}(h_{e}(A))-F^{i}_{cd}(A)$ is
a function of $A^{m}_{a}$ and its derivatives up to order $n$.  (If the
Hamiltonian is classical, we have $f_{cd}^i=0$.) It is obtained by expanding
the corresponding function $\tilde{F}^{i}_{cd}(h_{e}(A))$ that should appear
in place of the classical $F^{i}_{cd}(A)$ in an effective Hamiltonian computed
for a loop-quantized operator. There may also be a dependence on $E^a_i$ and
its spatial derivatives if there is lattice refinement
\cite{InhomLattice,CosConst}, in which case properties of the curves $e$ used
to construct a quantum Hamiltonian would depend on the spatial geometry. For
simplicity, we ignore such a dependence for a first analysis.

It is sufficient to assume that the holonomy-modification function
$f^{i}_{cd}(A^{m}_{a},\epsilon)$ is an antisymmetric tensor, just as
$F^{i}_{cd}$, because it is contracted with the antisymmetric combination
$\epsilon^{jk}{}_i E^c_jE^d_k$ of triad components.  We write the modified
Hamiltonian constraint as
\begin{eqnarray}
\label{EuclideanHhC}
H_Q[N]=\frac{1}{16\pi G}\int _{\Sigma}{\rm
  d}^{3}xN(\mathcal{H}+\delta \mathcal{H}_Q)=H[N]+\delta H_Q[N]\,.
\end{eqnarray}

After this modification, motivated by full loop quantum gravity, we may
perturb the Hamiltonian in order to describe cosmological inhomogeneity. We
use the splittings into background and inhomogeneity as given in
\cite{ConstraintAlgebra} (see also \cite{DeformedCosmo}).  Considering
perturbations around a spatially flat, homogeneous and isotropic metric, the
connection variables $A^{i}_{a}$ and the densitized triad $E^{a}_{i}$ can be
expanded as
\begin{eqnarray} \label{BPcurvature1}
&&
A^{i}_{a}=\bar{A}^{i}_{a}+\delta A^{i}_{a}:=\bar{q}\delta^{i}_{a}+\delta A^{i}_{a}
\\
&&
\label{BPEaiExpansion}
E^{a}_{i}=\bar{E}^{a}_{i}+\delta E^{a}_{i}:=\bar{p}\delta^{a}_{i}+\delta E^{a}_{i}
\end{eqnarray}
where the homogeneous mode is defined by
\begin{equation} \label{Hmode}
\bar{q}:=\frac{1}{3V_{0}}\int_{\Sigma} A^{i}_{a}\delta^{a}_{i}{\rm d}^{3}x
\quad ,\quad \bar{p}:=\frac{1}{3V_{0}}\int_{\Sigma}
E^{a}_{i}\delta^{i}_{a}{\rm d}^{3}x
\end{equation}
with $V_{0}=\int_{\Sigma} {\rm d}^{3}x$ (integrated over some fixed region, or
all of space if it is compact). We will assume $\bar{p}>0$, fixing the spatial
orientation. In order to avoid over-counting the degrees of freedom, the
perturbations $\delta E^{a}_{i}$ and $\delta A^{i}_{a}$ do not have
homogeneous modes:
\begin{equation} \label{PHmode}
\int_{\Sigma} \delta E^{a}_{i}\delta^{i}_{a}{\rm d}^{3}x=0
\quad,\quad \int_{\Sigma} \delta A^{i}_{a}\delta^{a}_{i}{\rm d}^{3}x=0\,.
\end{equation}
Therefore, the Poisson brackets of the background and perturbed variables can
be constructed as
\begin{equation} \label{PB}
\{\bar{q},\bar{p}\}=\frac{8\pi G}{3V_{0}}
\quad,\quad
\{\delta A^{i}_{a}(x),\delta E^{b}_{j}(y)\}=8\pi
G\delta^{i}_{j}\delta^{b}_{a}\delta^{3}(x-y)\,.
\end{equation}

We note \cite{DeformedCosmo} that there is a single inhomogeneous perturbation
$\delta f$ for any field component $f$, instead of a whole tower
$\delta^{(1)}f$, $\delta^{(2)}f$ and so on, as often used for linear
perturbation equations at all orders. The latter decomposition would be
convenient when one tries to solve a given set of equations of motion. In our
context, however, we first need to derive consistent forms of equations of
motion using canonical methods, which requires a well-defined Poisson or
symplectic structure. Since linearized perturbations $\delta^{(1)}f$,
$\delta^{(2)}f$ and so on would not provide independent degrees of freedom,
one cannot define a Poisson structure for them. The decomposition
(\ref{BPcurvature1}), by contrast, gives a well-defined Poisson structure
(\ref{PB}).

The background variables of the lapse function and shift vector
can be chosen as
\begin{equation} \label{BV1}
\bar{N}=\sqrt{\bar{p}}
\end{equation}
for conformal background time, and
\begin{equation}
  \bar{N}^{a}=0
\end{equation}
for an isotropic background.  Moreover, the perturbed lapse $\delta N$ does
not have homogeneous modes:
\begin{equation} \label{PHmode1}
\int_{\Sigma} \delta N {\rm d}^{3}x=0\,,
\end{equation}
just as (\ref{PHmode}).

In order to restrict attention to scalar modes, we shall parameterize the
basic perturbed phase space variables $(\delta A^{i}_{a},\delta E^{b}_{j})$ in
terms of suitable independent functions. As discussed in
\cite{ConstraintAlgebra}, $\delta E^{b}_{j}$ and the extrinsic-curvature
perturbation $\delta K^{i}_{a}$ can be parameterized as
\begin{eqnarray}\label{KEScalarModeParameter}
  \delta K^{i}_{a}=\delta^{i}_{a}\kappa_{1}+\partial_{a}\partial^{i}\kappa_{2}
\quad,\quad
\delta
E^{a}_{i}=\delta^{a}_{i}\varepsilon_{1}+\partial_{i}\partial^{a}\varepsilon_{2}
\end{eqnarray}
in terms of two pairs of scalar functions.  In addition, the spin connection
is
\begin{equation} \label{Spinconnection}
\Gamma^{i}_{a} = -\frac{1}{2} \epsilon^{ijk} E^{b}_{j}
\left( 2\partial_{[a} E^{k}_{b]} + E^{c}_{k} E^{l}_{a} \partial_{c} E^{l}_{b}
  - E^{k}_{a} \frac{\partial_{b}(\det E)}{\det E}\right)\,.
\end{equation}
Perturbing this equation at the linear level, one obtains
\begin{eqnarray}\label{SpinCScalarModeParameter}
\delta
\Gamma^{i}_{a}=\frac{1}{2\bar{p}}\epsilon^{\ ij}_{a}\partial_{j}
(\varepsilon_{1}+\Delta
\varepsilon_{2}) \,,
\end{eqnarray}
where $\Delta \varepsilon_{2}:=\partial_{a}\partial^{a} \varepsilon_{2}$.  The
connection variables $\delta A^{i}_{a}$ can therefore be expressed as
\begin{eqnarray}\label{AScalarModeParameter}
\delta A^{i}_{a}&=&\delta K^{i}_{a}+\delta \Gamma^{i}_{a}
\nonumber\\
&=&
\delta^{i}_{a}\kappa_{1}
+\partial_{a}\partial^{i}\kappa_{2}
+\frac{1}{2\bar{p}}\epsilon^{\ ij}_{a}\partial_{j}(\varepsilon_{1}+\Delta
\varepsilon_{2})\, .
\end{eqnarray}

It is easy to see that the Gauss constraint
\begin{eqnarray}
\label{LQGGC}
G[\Lambda] := \frac{1}{8\pi G\gamma}\int_{\Sigma}{\rm d}^{3}x \Lambda^{i}G_{i}
=\frac{1}{8\pi G\gamma}\int_{\Sigma}{\rm d}^{3}x
\Lambda^{i}(\partial_{a}E^{a}_{i}+\epsilon_{ij}^{\ \ k}A^{j}_{a}E^{a}_{k})
\end{eqnarray}
is automatically satisfied for the scalar modes.  However, there is still a
non-trivial gauge flow generated by the Gauss constraint, so that we will have
to make sure that all expressions are invariant under SU(2) transformations of
the connection and densitized triad. The latter can be done easily without
computing the extended brackets including the Gauss constraint. We will
therefore first focus on the brackets between Hamiltonian and diffeomorphism
constraints.

\section{Constraints}\label{ConstraintAlgebra}

We now perturb the constraints to second order in inhomogeneity, so that
non-trivial constraints are obtained which govern the gauge system of
linear perturbations. We will not restrict the inhomogeneity to scalar modes
right away, but only when doing so entails crucial simplifications.

\subsection{Perturbative constraints}

The diffeomorphism constraint of Euclidean general relativity can be expressed
as
\begin{eqnarray} \label{EGravityDiffeomorphismC}
D[N^{a}] &:=& \frac{1}{8\pi G}\int_{\Sigma}{\rm d}^{3}x N^{c}(-
F^{k}_{cd} E^{d}_{k})
\nonumber\\
&\approx& \frac{1}{8\pi G}\int_{\Sigma}{\rm d}^{3}x
N^{c}[(- \partial_{c}A^{k}_{d} + \partial_{d}A^{k}_{c})E^{d}_{k}
+A^{i}_{c}\partial_{a}E^{a}_{i}] \,,
\end{eqnarray}
where in the second line the Gauss constraint (\ref{LQGGC}) has been used.
Since (\ref{LQGGC}) vanishes for scalar modes, we do not need to distinguish
between the diffeomorphism and vector constraints, and either expression in
(\ref{EGravityDiffeomorphismC}) is good for our purposes.  Perturbing the first
expression (usually identified as the vector constraint), we have
\begin{eqnarray} \label{EGravityDiffeomorphismCPerturbationv1Scalar}
D[N^{a}] = \frac{1}{8\pi G}\int_{\Sigma}{\rm d}^{3}x \delta
N^{c}[-\bar{p} \partial_{c} (\delta^{d}_{k} \delta A^{k}_{d})
+ \bar{p} (\partial_{k} \delta A^{k}_{c})
- \bar{q}\bar{p}\epsilon^{d}_{\ cn}\delta
A^{n}_{d}-\bar{q}^{2}\epsilon^{k}_{\ cd}\delta E^{d}_{k}]\,.
\end{eqnarray}

The perturbative expression of the Hamiltonian density up to the second
order has been derived in the appendix of \cite{VectorSecond} as
$\mathcal{H}= \mathcal{H}^{(0)}+ \mathcal{H}^{(1)}+
\mathcal{H}^{(2)}$ with
\begin{eqnarray} \label{HamiltonianExpansion}
\mathcal{H}^{(0)} &=& 6\bar{q}^{2}\sqrt{\bar p},
\\
\mathcal{H}^{(1)} &=&
4 \bar{q}\sqrt{\bar{p}} \delta^c_j\delta A_c^j
+\frac{\bar{q}^{2}}{\sqrt{\bar{p}}} \delta_c^j\delta E^c_j
+2\sqrt{\bar{p}}\epsilon_{i}^{\ cd}\partial_{c}\delta A^{i}_{d},
\\
\mathcal{H}^{(2)} &=&
-\sqrt{\bar{p}} \delta A_c^j\delta A_d^k\delta^c_k\delta^d_j
+\sqrt{\bar{p}} (\delta A_c^j\delta^c_j)^2
+\frac{2\bar{q}}{\sqrt{\bar{p}}} \delta E^c_j\delta A_c^j
+\frac{\bar{q}^{2}}{2\bar{p}^{3/2}} \delta E^c_j\delta E^d_k\delta_c^k\delta_d^j
\nonumber\\
&& \quad
-\frac{\bar{q}^{2}}{4\bar{p}^{3/2}}(\delta E^c_j\delta_c^j)^2
+\frac{1}{\sqrt{\bar{p}}}\left( 4 \epsilon^{\ ck}_{i} \delta E^{d}_{k} -
  \epsilon^{\ cd}_{i} \delta E^{a}_{j} \delta^{j}_{a} \right) \partial_{[c}
\delta A^{i}_{d]}\, . \label{H2}
\end{eqnarray}
For a Hamiltonian constraint of the form (\ref{EuclideanHdwithHolonomy}), we
write
\begin{equation}
 f_{cd}^i= f^{i(0)}_{cd}+ f^{i(1)}_{cd}+ f^{i(2)}_{cd}
\end{equation}
expanded up to second order in inhomogeneity, and obtain the modification terms
\begin{eqnarray}
\delta \mathcal{H}_Q^{(0)} &=& \sqrt{\bar p}f^{i(0)}_{cd}\epsilon^{\ cd}_{i}~,
\\
\delta \mathcal{H}_Q^{(1)} &=&
\sqrt{\bar{p}}f^{i(1)}_{cd}\epsilon^{\ cd}_{i}
+\frac{f^{i(0)}_{cd}}{\sqrt{\bar{p}}} \left( 2 \epsilon^{\ ck}_{i} \delta E^{d}_{k} - \frac{1}{2} \epsilon^{\ cd}_{i} \delta E^{a}_{j} \delta^{j}_{a} \right)
,
\\
\delta \mathcal{H}_Q^{(2)} &=&
\sqrt{\bar{p}} f^{i(2)}_{cd}\epsilon^{\ cd}_{i}
+\frac{f^{i(1)}_{cd}}{\sqrt{\bar{p}}} \left( 2 \epsilon^{\ ck}_{i} \delta E^{d}_{k} - \frac{1}{2} \epsilon^{\ cd}_{i} \delta E^{a}_{j} \delta^{j}_{a} \right)
\nonumber\\
&& \quad
+\frac{f^{i(0)}_{cd}}{\bar{p}^{3/2}} \left[ \epsilon_{i}^{\ jk} \delta E^{c}_{j} \delta E^{d}_{k}
-\epsilon_{i}^{\ ck} \delta E^{d}_{k} \delta E^{a}_{j} \delta^{j}_{a}
+\frac{1}{8} \epsilon_{i}^{\ cd} (\delta E^{a}_{j} \delta^{j}_{a})^{2}
+\frac{1}{4} \epsilon_{i}^{\ cd} \delta E^{a}_{j} \delta E^{b}_{k}
\delta^{j}_{b} \delta^{k}_{a} \right] \,.
\end{eqnarray}
For later convenience, we denote
$\mathcal{F}^{(0)}=f^{i(0)}_{cd}\epsilon^{\ cd}_{i}$,
$\mathcal{F}^{(1)}=f^{i(1)}_{cd}\epsilon^{\ cd}_{i}$ and
$\mathcal{F}^{(2)}=f^{i(2)}_{cd}\epsilon^{\ cd}_{i}$.

At this stage, we pause and compare the parameterization with the one used in
\cite{ScalarHolInv} and related work. In these papers, $\delta K_a^i$ was used
instead of $\delta A_a^i$, and the derivative term present in the classical
constraint (\ref{H2}) could be eliminated using the Gauss constraint. The
Hamiltonian constraint then contains no derivatives of the field conjugate to
$\delta E^a_i$. However, even if such terms can be eliminated from the
classical Hamiltonian, they may appear in an effective constraint with a
derivative expansion of non-local holonomy modifications. Here, we assume that
they may be induced via the terms $f^{i(1)}_{cd}$ and $f^{i(2)}_{cd}$, up to a
certain order in derivatives.

\subsection{Brackets}

For computational purposes, it is convenient to split the perturbed
Hamiltonian and its modification terms into two parts each,
\begin{eqnarray}
\label{EuclideanPHCScalar}
&&
H[N]=\frac{1}{16\pi G}\int {\rm d}^{3}x
N\mathcal{H}=H[\bar{N}]+H[\delta N]\,,
\\
&&
\label{EuclideanPHCholonomyScalar}
\delta H_Q[N]=\frac{1}{16\pi G}\int {\rm d}^{3}x
N\delta \mathcal{H}_Q=\delta H_Q[\bar{N}]+\delta H_Q[\delta N]\,.
\end{eqnarray}
According to Eqs.~(\ref{PHmode}) and (\ref{PHmode1}), the integrals
$\int_{\Sigma} {\rm d}^{3}x \bar{N} \mathcal{H}^{(1)}$,
$\int_{\Sigma} {\rm d}^{3}x \delta N \mathcal{H}^{(0)}$,
$\int_{\Sigma} {\rm d}^{3}x \bar{N} \delta \mathcal{H}_Q^{(1)}$ and
$\int_{\Sigma} {\rm d}^{3}x \delta N \delta \mathcal{H}_Q^{(0)}$ are zero.
Therefore, the explicit expressions for the perturbed Hamiltonian constraint
are \cite{ConstraintAlgebra}
\begin{eqnarray}
\label{EuclideanPHCNbarScalar}
&&
H[\bar{N}]=\frac{1}{16\pi G}\int {\rm
  d}^{3}x\bar{N}[\mathcal{H}^{(0)}+\mathcal{H}^{(2)}]\quad ,\quad
H[\delta N]=\frac{1}{16\pi G}\int {\rm d}^{3}x\delta
N\mathcal{H}^{(1)}\,,
\label{EuclideanCPHC1Scalar}
\\
&&
\delta H_Q[\bar{N}]
= \frac{1}{16\pi G}\int {\rm
  d}^{3}x\bar{N}[\delta \mathcal{H}_Q^{(0)}+\delta \mathcal{H}_Q^{(2)}]\quad,\quad
\delta H_Q[\delta N]=\frac{1}{16\pi G}\int {\rm d}^{3}x\delta
N\delta \mathcal{H}_Q^{(1)}\,.
\end{eqnarray}

In a perturbative treatment, one may fix the background gauge so that
$H[\bar{N}]$ would generate equations of motion of background and perturbation
variables, while $H[\delta N]$ generates gauge transformations for the
modes. However, for consistency in the form of a closed set of
gauge-invariant observables, the constraints must be preserved by
evolution. Both types of generators must then come from a closed bracket of
constraints $H[\bar{N}+\delta N]$ together with $D[N^a]$.  As we have the
explicit expression for the perturbed Hamiltonian constraint at hand, we can
calculate the Poisson brackets between Hamiltonian and diffeomorphism
constraints and between two Hamiltonian constraints, and check whether they
can be closed.

Before proceeding, we shall assume that the
holonomy-modification function $f^{i}_{cd}$ is a function of the connection
variable $A^{m}_{a}$ up to first-order derivative, that is $f^{i}_{cd}\equiv
f^{i}_{cd}(A,\partial A,\epsilon)$, as used for vector modes in
\cite{VectorSecond}. Higher spatial derivatives require a more-involved
treatment by a systematic expansion as developed and applied to spherically
symmetric systems in \cite{HigherSpatial}. Here we assume the classical
derivative order but allow for all coefficients to be modified, thereby
extending the treatment of \cite{ScalarHolInv}.
In this case, the holonomy-modification function can be expanded as
\begin{eqnarray} \label{HCFExpansionCaseI}
&&f^{i}_{cd}(A,\partial A,\epsilon)
\nonumber\\
&=& \left.f^{i}_{cd}(A,\partial A,\epsilon)\right|_{\bar{A}^m_a}
+\left.\frac{\partial f^{i}_{cd}(A,\partial A,\epsilon)}{\partial
    A^{m}_{a}}\right|_{\bar{A}^{m}_{a}} \delta A^{m}_{a}
+\left.\frac{\partial f^{i}_{cd}(A,\partial A,\epsilon)}{\partial
    (\partial_{e} A^{m}_{a})}\right|_{\bar{A}^{m}_{a}}
\partial_{e} \delta A^{m}_{a}\nonumber\\
&&+\frac{1}{2}\left.\frac{\partial^{2} f^{i}_{cd}(A,\partial
    A,\epsilon)}{\partial A^{m}_{a} \partial
    A^{n}_{b}}\right|_{\bar{A}^{m}_{a}} \delta A^{m}_{a} \delta A^{n}_{b}
+\left.\frac{\partial^{2} f^{i}_{cd}(A,\partial A,\epsilon)}{\partial
    A^{m}_{a} \partial (\partial_{e} A^{n}_{b})}\right|_{\bar{A}^{m}_{a}}
\delta A^{m}_{a} \partial_{e}\delta A^{n}_{b}
\nonumber\\
&&
+\frac{1}{2}\left.\frac{\partial^{2} f^{i}_{cd}(A,\partial
    A,\epsilon)}{\partial (\partial_{e}A^{m}_{a}) \partial (\partial_{f}
    A^{n}_{b})}\right|_{\bar{A}^{m}_{a}} \partial_{e}\delta
A^{m}_{a} \partial_{f}\delta A^{n}_{b}
+\cdots
\nonumber\\
&=&f^{i(0)}_{cd}(\bar{q},\epsilon)
+\mathcal{A}^{i(1)}_{cd}(\bar{q},\delta A,\epsilon)
+\mathcal{B}^{i(1)}_{cd}(\bar{q},\partial\delta A,\epsilon)
+\mathcal{A}^{i(2)}_{cd}(\bar{q},\delta A,\epsilon)
+\mathcal{B}^{i(2)}_{cd}(\bar{q},\delta A,\partial\delta A,\epsilon)
\nonumber\\
&&
+\mathcal{C}^{i(2)}_{cd}(\bar{q},\partial\delta A,\epsilon)
+\cdots
\end{eqnarray}
For later convenience, we have denoted
$f^{i(1)}_{cd}\equiv\mathcal{A}^{i(1)}_{cd} +\mathcal{B}^{i(1)}_{cd}$ and
$f^{i(2)}_{cd}\equiv\mathcal{A}^{i(2)}_{cd}+\mathcal{B}^{i(2)}_{cd}
+\mathcal{C}^{i(2)}_{cd}$, where superscripts indicate orders of
inhomogeneity, and $\mathcal{A}$, $\mathcal{B}$, $\mathcal{C}$ derivative
orders.

Since we expand the Hamiltonian and diffeomorphism constraints up to second
order in inhomogeneity, higher-order terms in a power-series expansion by
$A_a^i$ of the holonomy-modification function will not provide independent
contributions of products of $\delta A_a^i$ but just modify the background
dependence of coefficients included here.  Therefore, it is enough to consider
the holonomy-modification function up to the second order in inhomogeneity,
even if it may come from non-polynomial functions such as the sine used in the
usual background modification. As already stated, our only assumption is that
no spatial derivatives of $A_a^i$ of orders higher than the classical one
appear.

We first consider the Poisson bracket between Hamiltonian and diffeomorphism
constraints,
\begin{eqnarray}
\label{HDScalar}
\{H[N],D[N^{a}]\}=\{H[\bar{N}],D[N^{a}]\}+\{H[\delta
N],D[N^{a}]\}\,.
\end{eqnarray}
It is straightforward to show that the Poisson bracket
$\{H[\bar{N}],D[N^{a}]\}$ vanishes, and hence we have
\begin{eqnarray}
\label{HDdeltaNNaScalar}
\{H[N],D[N^{a}]\}=\{H[\delta N],D[N^{a}]\}
=H[\delta N^{c}\partial_{c}\delta N]\,.
\end{eqnarray}

Note that in Euclidean signature one commonly employs the diffeomorphism
constraint with a sign opposite to that in Lorentzian general relativity, so
that there is a sign difference between the above Poisson bracket and
corresponding one in Lorentzian signature. There are similar results in the
following Poisson brackets, including the classical case and that with
holonomy modifications.  Thus the Poisson bracket between perturbed classical
Hamiltonian and diffeomorphism constraints agrees with the bracket between the
original classical constraints. This indicates the consistency between the
perturbed constraint expressions and elementary Poisson brackets including the
background and perturbed basic variables.

We shall now derive the Poisson bracket between the Hamiltonian and the
diffeomorphism constraints when the former includes holonomy modifications.
It should be noted that for vector modes in \cite{VectorSecond}, there is no
lapse perturbation, that is $\delta N=0$, and $\delta H_Q[\delta N]$
vanishes. But for scalar modes we have $\delta N\neq 0$, so that we need to
calculate both Poisson brackets, $\{\delta H_Q[\bar{N}],D[N^{a}]\}$ and
$\{\delta H_Q[\delta N],D[N^{a}]\}$.

We calculate the first Poisson bracket:
\begin{eqnarray} \label{HQNbarDScalar}
&&\{\delta H_Q[\bar{N}],D[N^{a}]\}
\nonumber\\
&&
= \frac{1}{16\pi G}\int {\rm d}^{3}x \delta N^{c}
\Bigl[-\frac{1}{2}\bar{q}\delta^{i}_{c}\frac{\partial \mathcal{A}^{(1)}}{\partial (\delta A^{i}_{a})}\partial_{a}(\delta E^{d}_{k}\delta^{k}_{d})
+f^{i(0)}_{bc}\epsilon^{\ bj}_{i}\partial_{j}(\delta E^{d}_{k}\delta^{k}_{d})
-\frac{2}{3}\mathcal{F}^{(0)}\delta^{k}_{c}(\partial_{d}\delta E^{d}_{k})
\nonumber\\
&&
-2f^{i(0)}_{cd} \epsilon^{\ jk}_{i} \partial_{j} \delta E^{d}_{k}
+2\bar{q} \frac{\partial \mathcal{A}^{j(1)}_{bd}}{\partial(\delta A^{i}_{a})}\epsilon^{\ bk}_{j}\delta^{i}_{c}\partial_{a}\delta E^{d}_{k}
+\frac{\bar{q}}{2}\frac{\partial \mathcal{B}^{(1)}}{\partial(\partial_{e}\delta A^{i}_{a})}\delta^{i}_{c}\partial_{a}\partial_{e}(\delta E^{d}_{k}\delta^{k}_{d})
\nonumber\\
&&
-2\bar{q} \frac{\partial \mathcal{B}^{j(1)}_{bd}}{\partial (\partial_{e}\delta A^{i}_{a})}
\epsilon^{\ bk}_{j}\delta^{i}_{c}\partial_{a}\partial_{e}\delta E^{d}_{k}
+\frac{1}{3}\bar{p}\frac{\partial \mathcal{F}^{(0)}}{\partial \bar{q}}\partial_{k} \delta A^{k}_{c}
+\bar{q}\bar{p}\delta^{i}_{c}\partial_{a}\frac{\partial \mathcal{A}^{(2)}}{\partial (\delta A^{i}_{a})}
+\bar{q}\bar{p}\delta^{i}_{c}\partial_{a}\frac{\partial \mathcal{B}^{(2)}}{\partial (\delta A^{i}_{a})}
\nonumber\\
&&
-\bar{q}\bar{p}\delta^{i}_{c}\partial_{a}\partial_{e}\frac{\partial
  \mathcal{B}^{(2)}}{\partial (\partial_{e}\delta A^{i}_{a})}
-\bar{q}\bar{p}\delta^{i}_{c}\partial_{a}\partial_{e}\frac{\partial
  \mathcal{C}^{(2)}}{\partial (\partial_{e}\delta A^{i}_{a})}
+\bar{p}\partial_{c}\mathcal{A}^{(1)}
+\bar{p}\partial_{c}\mathcal{B}^{(1)}
\nonumber\\
&&
-2\bar{p} \epsilon^{\ bj}_{i}\partial_{j}\mathcal{A}^{i(1)}_{bc}
-2\bar{p} \epsilon^{\ bj}_{i}\partial_{j}\mathcal{B}^{i(1)}_{bc}
-\frac{\bar{p}}{3}\frac{\partial \mathcal{F}^{(0)}}{\partial
  \bar{q}}\partial_{c}(\delta^{d}_{k}\delta A^{k}_{d})\Bigr]\,.
\end{eqnarray}
Hence in contrast to the classical case, the Poisson bracket
$\{H_Q[\bar{N}],D[N^{a}]\}$ does not vanish identically due to the
introduction of holonomy effects.  The second
Poisson bracket is
\begin{eqnarray} \label{HQdeltaNDScalar}
&&\{\delta H_Q[\delta N],D[N^{a}]\}
\nonumber\\
&&
=\frac{1}{16\pi G}\int {\rm d}^{3}x \Bigl[(\delta N^{i}\partial_{a}\delta
  N)
\left(\bar{q}\sqrt{\bar{p}}\frac{\partial \mathcal{A}^{(1)}}{\partial(\delta
    A^{i}_{a})}
-2\sqrt{\bar{p}}f^{k(0)}_{di}\epsilon^{\ da}_{k}\right)
\nonumber\\
&&
+(\delta N^{c}\partial_{c}\delta N)
\sqrt{\bar{p}}\mathcal{F}^{(0)}
-(\delta N^{i}\partial_{a}\partial_{e}\delta N)\bar{q}\sqrt{\bar{p}}\frac{\partial \mathcal{B}^{(1)}}{\partial(\partial_{e}\delta A^{i}_{a})}\Bigr]\,.
\end{eqnarray}
The Poisson bracket we are looking for is the sum of
Eqs.~(\ref{HDdeltaNNaScalar}), (\ref{HQNbarDScalar}) and
(\ref{HQdeltaNDScalar}),
\begin{eqnarray} \label{HQNDScalar}
\{H_Q[N],D[N^{a}]\}=\{H[\delta N],D[N^a]\}+\{H_Q[\bar{N}],D[N^{a}]\}+\{H_Q[\delta
N],D[N^{a}]\} \,.
\end{eqnarray}
We will discuss possible anomaly-free versions in the next section.

We now calculate the Poisson bracket between two Hamiltonian constraints,
smeared with different functions $N_{1}= \bar{N}+\delta N_1$ and $N_{2}=
\bar{N}+\delta N_2$. We have $\{H[\delta N_{1}],H[\delta N_{2}]\}=0$ because
the absence of a background term in the diffeomorphism constraint implies that
the leading non-zero contribution would be of third order, which is eliminated
in our second-order expansion.  We therefore have
\begin{eqnarray}
\label{PBtwoHCClassicalScalar}
\{H[N_{1}],H[N_{2}]\}
&=& \{H[\delta N_1],H[\bar{N}]\}+ \{H[\bar{N}],H[\delta
N_2]\} \nonumber\\
&=&\{H[\delta N_{1}-\delta N_{2}],H[\bar{N}]\}
=-D\left[\frac{\bar{N}}{\bar{p}}\partial^{c}(\delta N_{2}-\delta
  N_{1})\right] \,.
\end{eqnarray}
Again, Eq.~(\ref{PBtwoHCClassicalScalar}) confirms the consistency between the
perturbed constraint expressions and elementary Poisson brackets including the
background and perturbed basic variables.

With holonomy modifications, we similarly have
\begin{eqnarray}\label{PBC}
\{H_Q[N_{1}],H_Q[N_{2}]\}&=& \{H_Q[\delta N_{1}-\delta N_{2}],H_Q[\bar{N}]\}
\nonumber\\
&=& \{H[\delta N_1-\delta N_2],H[\bar{N}]\}+ \{H[\delta N_1-\delta N_2],\delta
H_Q[\bar{N}]\}\nonumber\\
&&+ \{\delta H_Q[\delta N_1-\delta N_2],H[\bar{N}]\}+ \{\delta
H_Q[\delta N_1-\delta N_2],\delta H_Q[\bar{N}]\}
\end{eqnarray}
where
\begin{eqnarray}\label{PBCQ}
&&
\{H[\delta N_{1}-\delta N_{2}],\delta H_Q[\bar{N}]\}
\nonumber\\
&=&
\frac{1}{8\pi G}\int {\rm d}^{3}x (\delta N_{1}-\delta N_{2})
\Bigl\{\Bigl[\frac{\bar{q}}{24\sqrt{\bar{p}}} \left(8\mathcal{F}^{(0)}+\bar{q}\frac{\partial \mathcal{F}^{(0)}}{\partial \bar{q}}\right)
+\frac{\bar{q}^2}{8\sqrt{\bar{p}}}\frac{\partial\mathcal{A}^{(1)}}{\partial (\delta A^{i}_{a})}\delta^{i}_{a}\Bigr]
\delta E^{d}_{k}\delta^{k}_{d}
\nonumber\\
&&
-\Bigl[
\frac{\bar{q}}{\sqrt{\bar{p}}} f^{i(0)}_{cd}\epsilon^{\ ck}_{i}
+\frac{\bar{q}^2}{2\sqrt{\bar{p}}}\frac{\partial
\mathcal{A}^{j(1)}_{cd}}{\partial (\delta A^{i}_{a})}
\delta^i_a
\epsilon^{\ ck}_{j}
\Bigr]\delta E^{d}_{k}
+\frac{\sqrt{\bar{p}}}{6}
\left(\mathcal{F}^{(0)}-\bar{q}\frac{\partial \mathcal{F}^{(0)}}{\partial \bar{q}}\right)
\delta A^{k}_{d}\delta^{d}_{k}
\nonumber\\
&&
+\frac{1}{2}\bar{q}\sqrt{\bar{p}}\mathcal{F}^{(1)}
-\frac{\bar{q}^2\sqrt{\bar{p}}}{4}\delta^i_a
\left(\frac{\partial\mathcal{A}^{(2)}}{\partial (\delta A^{i}_{a})}
+\frac{\partial\mathcal{B}^{(2)}}{\partial (\delta A^{i}_{a})}\right)
\Bigr\}
\nonumber\\
&&
+\frac{1}{8\pi G}\int {\rm d}^{3}x \partial_{e}(\delta N_{1}- \delta N_{2})
\Bigl\{
\Bigl[
\frac{1}{2\sqrt{\bar{p}}}f^{i(0)}_{cb}\epsilon^{\ cj}_{i}\epsilon^{\ eb}_{j}
+\frac{\bar{q}^2}{8\sqrt{\bar{p}}}\frac{\partial \mathcal{B}^{(1)}}{\partial (\partial_{e}\delta A^{i}_{a})}\delta^i_a
\Bigr]
\delta E^{d}_{k}\delta^{k}_{d}
\nonumber\\
&&
+
\frac{1}{\sqrt{\bar{p}}}
\Bigl[-\frac{ \mathcal{F}^{(0)}}{4}
-f^{i(0)}_{bd}\epsilon^{\ jk}_{i}\epsilon^{\ eb}_{j}
+\frac{\bar{q}^2}{2}\delta^{i}_{a}
\epsilon^{\ ck}_{j}
\frac{\partial \mathcal{B}^{j(1)}_{cd}}{\partial (\partial_{e}\delta A^{i}_{a})}
\Bigr] \delta E^{d}_{k}
+\frac{\sqrt{\bar{p}}}{12}
\frac{\partial \mathcal{F}^{(0)}}{\partial \bar{q}}
\epsilon^{\ ed}_{i}
\delta A^i_d
\nonumber\\
&&
-f^{i(1)}_{db}\epsilon^{\ dj}_{i}\epsilon^{\ eb}_{j}
+\frac{\sqrt{\bar{p}}\bar{q}^2}{4}
\left(\frac{\partial \mathcal{B}^{(2)}}{\partial (\partial_{e}\delta A^{i}_{a})}
+\frac{\partial \mathcal{C}^{(2)}}{\partial (\partial_{e}\delta A^{i}_{a})}\right)\delta^i_a
\Bigr\}
\end{eqnarray}
for the first non-classical bracket,
\begin{eqnarray}\label{PBQC}
&&
\{\delta H_Q[\delta N_{1}-\delta N_{2}],H[\bar{N}]\}
\nonumber\\
&=&
\frac{1}{8\pi G}\int {\rm d}^{3}x (\delta N_{1}-\delta N_{2})
\Bigl\{-\frac{\bar{q}^2}{8\sqrt{\bar{p}}}
\left(\frac{\partial \mathcal{F}^{(0)}}{\partial \bar{q}}
+\frac{\partial \mathcal{A}^{(1)}}{\partial (\delta A^i_a)}\delta^i_a\right)
\delta E^{d}_{k}\delta^{k}_{d}
-\frac{\sqrt{\bar{p}}}{2}\mathcal{F}^{(0)}\delta A^k_d\delta^d_k
\nonumber\\
&&
+\Bigl[
\frac{\bar{q}}{\sqrt{\bar{p}}}
\left(f^{i(0)}_{cd}+\frac{\bar{q}}{4}\frac{\partial f^{i(0)}_{cd}}{\partial \bar{q}}\right)
+\frac{\bar{q}^2}{4\sqrt{\bar{p}}}
\frac{\partial \mathcal{A}^{(1)}}{\partial (\delta A^{i}_{a})}
\delta^k_a\delta^i_d
\Bigr]\delta E^{d}_{k}
+\left(\frac{1}{2}\bar{q}\sqrt{\bar{p}}
\frac{\partial\mathcal{A}^{(1)}}{\partial(\delta A^i_a)}
+\sqrt{\bar{p}}f^{j(0)}_{ca}\epsilon^{\ cd}_j\delta^a_k
\right)\delta A^k_d
\nonumber\\
&&
+\frac{\bar{q}\sqrt{\bar{p}}}{4}
\left(\bar{q}\frac{\partial\mathcal{F}^{(1)}}{\partial\bar{q}}
-2\mathcal{F}^{(1)}\right)
\Bigr\}
\nonumber\\
&&
+\frac{1}{8\pi G}\int {\rm d}^{3}x \partial_{e}(\delta N_{1}- \delta N_{2})
\Bigl\{
\Bigl[\frac{\bar{q}^2}{8\sqrt{\bar{p}}}
\frac{\partial\mathcal{B}^{(1)}}{\partial (\partial_{e}\delta A^{i}_{a})}
\delta^i_a
+\frac{1}{2\sqrt{\bar{p}}}
f^{j(0)}_{ba}\epsilon^{\ bi}_j\epsilon^{\ ea}_i
\Bigr]\delta E^d_k\delta^k_d
\nonumber\\
&&
+\Bigl[-\frac{\bar{q}^2}{4\sqrt{\bar{p}}}
\frac{\partial\mathcal{B}^{(1)}}{\partial (\partial_{e}\delta A^{i}_{a})}
\delta^k_a\delta^i_d
+\frac{1}{4\sqrt{\bar{p}}}\mathcal{F}^{(0)}\epsilon^{\ ek}_d
-\frac{1}{\sqrt{p}}f^{j(0)}_{bd}\epsilon^{\ bi}_j\epsilon^{\ ek}_i
+\frac{1}{\sqrt{p}}f^{j(0)}_{ba}\epsilon^{\ bi}_j\epsilon^{\ ak}_i\delta^e_d
\Bigr]\delta E^d_k
\nonumber\\
&&
+\frac{1}{4\sqrt{\bar{p}}}
\frac{\partial\mathcal{A}^{(1)}}{\partial \delta A^{i}_{a}}
\epsilon^{\ ed}_a\delta A^i_d
+\frac{1}{4\sqrt{\bar{p}}}
\frac{\partial\mathcal{B}^{(1)}}{\partial(\partial_e \delta A^{i}_{a})}
\epsilon^{\ cd}_a\partial_c\delta A^i_d
\Bigr\}
\end{eqnarray}
for the second non-classical bracket, and
\begin{eqnarray}
\label{PBtwoHCQScalar}
&&\{\delta H_Q[\delta N_{1}-\delta N_{2}],\delta H_Q[\bar{N}]\}
\nonumber\\
&=&
\frac{1}{8\pi G}\int {\rm d}^{3}x (\delta N_{1}-\delta N_{2})
\Bigl\{-\frac{1}{24\sqrt{\bar{p}}}\mathcal{F}^{(0)}\frac{\partial \mathcal{F}^{(0)}}{\partial \bar{q}}
\delta E^{d}_{k}\delta^{k}_{d}
+\Bigl[\frac{1}{12\sqrt{\bar{p}}} \mathcal{F}^{(0)} \frac{\partial f^{i(0)}_{cd}}{\partial \bar{q}} \epsilon^{\ ck}_{i}
+\frac{1}{12\sqrt{\bar{p}}} f^{i(0)}_{cd} \frac{\partial \mathcal{F}^{(0)}}{\partial \bar{q}} \epsilon^{\ ck}_{i}
\nonumber\\
&&
+\frac{1}{8\sqrt{\bar{p}}} \mathcal{F}^{(0)} \frac{\partial \mathcal{A}^{(1)}}{\partial (\delta A^{i}_{a})}\delta^{i}_{d}\delta^{k}_{a}
-\frac{1}{4\sqrt{\bar{p}}} f^{j(0)}_{cd} \epsilon^{\ ck}_{j} \frac{\partial \mathcal{A}^{(1)}}{\partial (\delta A^{i}_{a})}\delta^{i}_{a}
+\frac{1}{2\sqrt{\bar{p}}} f^{j(0)}_{ad} \epsilon^{\ ik}_{j} \frac{\partial \mathcal{A}^{(1)}}{\partial (\delta A^{i}_{a})}\nonumber\\
&&-\frac{1}{2\sqrt{\bar{p}}}\left(-\frac{1}{2}\mathcal{F}^{(0)}\delta^{i}_{a}
+2f^{k(0)}_{ca}\epsilon^{\ ci}_{k}\right)\frac{\partial
\mathcal{A}^{j(1)}_{cd}}{\partial (\delta A^{i}_{a})}\epsilon^{\ ck}_{j}
\Bigr]\delta E^{d}_{k}
+\frac{\sqrt{\bar{p}}}{24} \left(\mathcal{F}^{(0)}\frac{\partial \mathcal{F}^{(1)}}{\partial \bar{q}}
-\mathcal{F}^{(1)}\frac{\partial \mathcal{F}^{(0)}}{\partial \bar{q}}\right)
\nonumber\\
&&-\frac{\sqrt{\bar{p}}}{8} \mathcal{F}^{(1)} \frac{\partial \mathcal{A}^{(1)}}{\partial (\delta A^{i}_{a})} \delta^{i}_{a}
+\frac{\sqrt{\bar{p}}}{2} f^{k(1)}_{ca} \epsilon^{\ ci}_{k} \frac{\partial \mathcal{A}^{(1)}}{\partial (\delta A^{i}_{a})}
-\frac{\sqrt{\bar{p}}}{4} \left(-\frac{1}{2}\mathcal{F}^{(0)}\delta^{i}_{a}
+2f^{k(0)}_{ca}\epsilon^{\ ci}_{k}\right) \left(\frac{\partial
  \mathcal{A}^{(2)}}{\partial (\delta A^{i}_{a})}+\frac{\partial
  \mathcal{B}^{(2)}}{\partial (\delta A^{i}_{a})}\right)
\Bigr\}
\nonumber\\
&&
+\frac{1}{8\pi G}\int {\rm d}^{3}x \partial_{e}(\delta N_{1}- \delta N_{2})
\Bigl\{
\frac{\sqrt{\bar{p}}}{4} \left(-\frac{1}{2}\mathcal{F}^{(0)}\delta^{i}_{a}
+2f^{k(0)}_{ca}\epsilon^{\ ci}_{k}\right)  \frac{\partial
\mathcal{B}^{(1)}}{\partial (\partial_{e}\delta A^{i}_{a})}
\nonumber\\
&&
+
\Bigl[-\frac{1}{8\sqrt{\bar{p}}} \mathcal{F}^{(0)}
\frac{\partial \mathcal{B}^{(1)}}{\partial (\partial_{e}\delta A^{i}_{a})} \delta^{i}_{d} \delta^{k}_{a}
+\frac{1}{4\sqrt{\bar{p}}} f^{j(0)}_{cd}\epsilon^{\ ck}_{j} \frac{\partial \mathcal{B}^{(1)}}{\partial (\partial_{e}\delta A^{i}_{a})} \delta^{i}_{a}
-\frac{1}{2\sqrt{\bar{p}}} f^{j(0)}_{ad}\epsilon^{\ ik}_{j}
\frac{\partial \mathcal{B}^{(1)}}{\partial (\partial_{e}\delta A^{i}_{a})}
\nonumber\\
&&
-\frac{1}{2\sqrt{\bar{p}}} \left(-\frac{1}{2}\mathcal{F}^{(0)}\delta^{i}_{a}
+2f^{l(0)}_{ba}\epsilon^{\ bi}_{l}\right) \epsilon^{\ ck}_{j} \frac{\partial \mathcal{B}^{j(1)}_{cd}}{\partial (\partial_{e}\delta A^{i}_{a})}
\Bigr] \delta E^{d}_{k}
+\frac{\sqrt{\bar{p}}}{8} \mathcal{F}^{(1)}\delta^{i}_{a} \frac{\partial \mathcal{B}^{(1)}}{\partial (\partial_{e}\delta A^{i}_{a})}
\nonumber\\
&&
-\frac{\sqrt{\bar{p}}}{2} f^{k(1)}_{ca} \epsilon^{\ ci}_{k} \frac{\partial \mathcal{B}^{(1)}}{\partial (\partial_{e}\delta A^{i}_{a})}
- \frac{\sqrt{\bar{p}}}{4}
\left(-\frac{1}{2}\mathcal{F}^{(0)}\delta^{i}_{a}+2f^{l(0)}_{ba}\epsilon^{\ bi}_{l}\right)
\left(\frac{\partial \mathcal{B}^{(2)}}{\partial (\partial_{e}\delta A^{i}_{a})}
+\frac{\partial \mathcal{C}^{(2)}}{\partial (\partial_{e}\delta A^{i}_{a})}\right)
\Bigr\}
\end{eqnarray}
for the last bracket.

\section{Holonomy modification function and anomaly freedom}\label{Form1}

We can now check whether there are some specific forms of the holonomy
modification function which imply that the constraints are anomaly free.
The general form \cite{VectorSecond}
\begin{eqnarray}\label{ficdconcrete}
f^{i}_{cd}&=&\sigma(\bar{q})\epsilon^{i}_{\ cd}
+\slashed{\sigma}(\bar{q})\epsilon^{i}_{\ cd} A^{j}_{a}\delta^{a}_{j}
+\mu(\bar{q}) A^{i}_{b}\epsilon^{b}_{\ cd}
+ \nu(\bar{q})(\epsilon^{i}_{\ md} A^{m}_{c} + \epsilon^{i}_{\
  cn}A^{n}_{d})
\nonumber\\
&&
+\widetilde{\slashed{\sigma}}(\bar{q})\epsilon^{i}_{\ cd}
(A^{j}_{a}\delta^{a}_{j})^{2}
+\slashed{\mu}(\bar{q}) A^{i}_{b}\epsilon^{b}_{\ cd}
A^{j}_{a}\delta^{a}_{j}
+\slashed{\nu}(\bar{q})(\epsilon^{i}_{\ md} A^{m}_{c} +
\epsilon^{i}_{\ cn} A^{n}_{d}) A^{j}_{a}\delta^{a}_{j}
\nonumber\\
&&
+ \rho(\bar{q}) \epsilon^{i}_{\ mn} A^{m}_{c} A^{n}_{d}
+ \tau(\bar{q})\partial_{[c} A^{i}_{d]}\,,
\end{eqnarray}
of holonomy modification functions satisfies our previous assumptions:
antisymmetry in $c$ and $d$ as well as up to first-order derivatives of
$A_a^i$. At this point, there is no term of the form $A_a^i\partial_b A_c^j$
because on shell $A_a^i$ appears as a first-order (time) derivative. The
omitted term would therefore be considered to be of second total derivative
order and should not be included in a first-order derivative
expansion. This treatment of derivatives has been shown to be consistent in
\cite{HigherSpatial}.  The dependence of coefficients on the background
connection $\bar{q}$ is unrestricted, as it may result from an expansion of a
non-quadratic function of the connection. However, if $f_{cd}^i$ is expected
to result from a function of holonomies, expanded up to $n$-th order in a
dependence on $A_a^i$, the perturbation expansion implies that
$\sigma(\bar{q})$ is a polynomial of the same degree $n$, while
$\slashed{\sigma}(\bar{q})$, $\mu(\bar{q})$, $\nu(\bar{q})$ and
$\tau(\bar{q})$ are polynomials of degree $n-1$, and
$\widetilde{\slashed{\sigma}}(\bar{q})$, $\slashed{\mu}(\bar{q})$,
$\slashed{\nu}(\bar{q})$ and $\rho(\bar{q})$ are polynomials of degree $n-2$.

In the notation of (\ref{HCFExpansionCaseI}), the expressions of the
holonomy modification function $f^{i}_{cd}$ up to first order can be found as
\begin{eqnarray} \label{ficdexpandCaseIScalar}
f^{i(0)}_{cd} &=& (\sigma + 3\slashed{\sigma} \bar{q} + \mu \bar{q} + 2 \nu \bar{q} + \rho\bar{q}^{2}
+9\widetilde{\slashed{\sigma}}\bar{q}^{2}
+3\slashed{\mu}\bar{q}^{2}+6\slashed{\nu}\bar{q}^{2}
)\epsilon^{i}_{\ cd}\,,
\nonumber\\
\mathcal{A}^{i(1)}_{cd} &=&
(\nu+\rho\bar{q}+3\slashed{\nu}\bar{q})
(\epsilon^{i}_{\ md}\delta A^{m}_{c}+\epsilon^{i}_{\ cn}\delta A^{n}_{d})
+(\slashed{\sigma}
+6\widetilde{\slashed{\sigma}}\bar{q}
+\slashed{\mu}\bar{q}
+2\slashed{\nu}\bar{q})\delta A^{j}_{a}\delta^{a}_{j}\epsilon^{i}_{\ cd}
\nonumber\\
&&
+(\mu+3\slashed{\mu}\bar{q})\delta A^{i}_{b}\epsilon^{b}_{\ cd}\,,
\nonumber\\
\mathcal{B}^{i(1)}_{cd} &=& \tau \partial_{[c}\delta A^{i}_{d]} \,,
\nonumber\\
\mathcal{A}^{i(2)}_{cd} &=&
\rho\epsilon^{i}_{\ mn}\delta A^{m}_{c} \delta A^{n}_{d}
+\widetilde{\slashed{\sigma}}(\delta A^{j}_{a}\delta^{a}_{j})^{2}\epsilon^{i}_{\ cd}
+\slashed{\nu}(\epsilon^{i}_{\ md}\delta A^{m}_{c}+\epsilon^{i}_{\ cn}\delta A^{n}_{d})\delta A^{j}_{a}\delta^{a}_{j}
+\slashed{\mu} \delta A^{i}_{b}\epsilon^{b}_{\ cd}\delta A^{j}_{a}\delta^{a}_{j}\,,
\nonumber\\
\mathcal{B}^{i(2)}_{cd}&=&0\,,
\nonumber\\
\mathcal{C}^{i(2)}_{cd}&=& 0\,.
\end{eqnarray}
Note that for vector modes, with $\delta^b_j\delta A^j_b=0$, these equations
reduce to Eq.~(35) in \cite{VectorSecond}.  We list the following relations
for later convenience,
\begin{eqnarray} \label{ficdexpandNecessaryRelationCaseIScalar}
\mathcal{F}^{(0)} &=& 6(\sigma + 3\slashed{\sigma} \bar{q} + \mu \bar{q} + 2
\nu \bar{q} + \rho\bar{q}^{2}
+9\widetilde{\slashed{\sigma}}\bar{q}^{2}
+3\slashed{\mu}\bar{q}^{2}+6\slashed{\nu}\bar{q}^{2}
)\,,
\nonumber\\
\mathcal{A}^{(1)} &=&
(4\nu+6\slashed{\sigma}+2\mu
+4\rho\bar{q}+24\slashed{\nu}\bar{q}+
36\widetilde{\slashed{\sigma}}\bar{q}+12\slashed{\mu}\bar{q}
)\delta A^{k}_{d}\delta^{d}_{k}\,,
\nonumber\\
\mathcal{A}^{(2)} &=&
(\rho+6\widetilde{\slashed{\sigma}}+2\slashed{\nu}+2\slashed{\mu})
(\delta A^{k}_{d}\delta^{d}_{k})^{2}
-\rho \delta A^{m}_{c}\delta A^{n}_{d}\delta^{c}_{n}\delta^{d}_{m}\,,
\nonumber\\
f^{i(0)}_{cd}&=&\frac{\mathcal{F}^{(0)}}{6}\epsilon^{i}_{\ cd}\,,
\nonumber\\
\mathcal{B}^{(1)} &=& \tau \partial_{c}(\delta A^{i}_{d}\epsilon^{\ cd}_{i})\,.
\end{eqnarray}

\subsection{Hamiltonian and diffeomorphism constraints}

Substituting Eq.~(\ref{ficdexpandCaseIScalar}) into the expression of the
Poisson bracket (\ref{HQNbarDScalar}) and using the relations in
Eq.~(\ref{ficdexpandNecessaryRelationCaseIScalar}), we have
\begin{eqnarray} \label{HQNbarDfCaseIScalar}
&&\{\delta H_Q[\bar{N}],D[N^{a}]\}
\nonumber\\
&&
= \frac{1}{16\pi G}\int {\rm d}^{3}x \delta N^{c}
\Bigl[-2\left(\frac{\sigma}{\bar{q}} + 3\slashed{\sigma} + 2\mu + \nu
+9\bar{q}\widetilde{\slashed{\sigma}}
+6\bar{q}\slashed{\mu}
+3\bar{q}\slashed{\nu}
\right)\bar{q}\delta^{k}_{c}(\partial_{d}\delta E^{d}_{k})
\nonumber\\
&&
+2\Biggl(\frac{\partial\sigma}{\partial \bar{q}}
+3\bar{q}\frac{\partial\slashed{\sigma}}{\partial \bar{q}}
+\bar{q} \frac{\partial\mu}{\partial \bar{q}}
+2\bar{q} \frac{\partial\nu}{\partial \bar{q}}
+3\bar{q}^{2}\frac{\partial\slashed{\mu}}{\partial \bar{q}}
+\bar{q}^{2}\frac{\partial\rho}{\partial \bar{q}}
+9\bar{q}^{2}\frac{\partial \widetilde{\slashed{\sigma}}}{\partial \bar{q}}
+6\bar{q}^{2}\frac{\partial \slashed{\nu}}{\partial \bar{q}}
\nonumber\\
&&
+3\slashed{\sigma}+2\mu+\nu+18\bar{q}\widetilde{\slashed{\sigma}}
+9\bar{q}\slashed{\mu}
+9\bar{q}\slashed{\nu}
\Biggr)\bar{p}\partial_{k}\delta A^{k}_{c}
\nonumber\\
&&
-2\Biggl(\frac{\partial\sigma}{\partial \bar{q}}
+3\bar{q}\frac{\partial\slashed{\sigma}}{\partial \bar{q}}
+\bar{q} \frac{\partial\mu}{\partial \bar{q}}
+2\bar{q} \frac{\partial\nu}{\partial \bar{q}}
+3\bar{q}^{2}\frac{\partial\slashed{\mu}}{\partial \bar{q}}
+\bar{q}^{2}\frac{\partial\rho}{\partial \bar{q}}
+9\bar{q}^{2}\frac{\partial \widetilde{\slashed{\sigma}}}{\partial \bar{q}}
+6\bar{q}^{2}\frac{\partial \slashed{\nu}}{\partial \bar{q}}
\nonumber\\
&&
+2\slashed{\sigma}+\mu+\nu
+3\bar{q}\slashed{\mu}
+5\bar{q}\slashed{\nu}
+6\bar{q}\tilde{\slashed{\sigma}}
\Biggr)\bar{p}\partial_{c}(\delta^{d}_{k}\delta A^{k}_{d})
\nonumber\\
&&
+(\slashed{\sigma}
+6\bar{q}\widetilde{\slashed{\sigma}}
+\mu
+4\bar{q}\slashed{\mu}
+2\bar{q}\slashed{\nu})\bar{q}\partial_{c}(\delta E^{d}_{k}\delta^{k}_{d})
\Bigr]\,.
\end{eqnarray}
For scalar modes (\ref{KEScalarModeParameter}) and
(\ref{AScalarModeParameter}), we have the relations
\begin{eqnarray} \label{SomeRelationScalarConcrete}
&&
\delta^{k}_{c}(\partial_{d}\delta E^{d}_{k})
=\partial_{c}\varepsilon_{1}
+\partial_{c}(\Delta\varepsilon_{2}),
\nonumber\\
&&
\partial_{c}(\delta E^{d}_{k}\delta^{k}_{d})
=3\partial_{c}\varepsilon_{1}
+\partial_{c}(\Delta\varepsilon_{2}),
\nonumber\\
&&
\partial_{k}\delta A^{k}_{c}
=\partial_{c}\kappa_{1}
+\partial_{c}(\Delta\kappa_{2}),
\nonumber\\
&&
\partial_{c}(\delta A^{k}_{d}\delta^{d}_{k})
=3\partial_{c}\kappa_{1}
+\partial_{c}(\Delta\kappa_{2}),
\nonumber\\
&&
\epsilon^{k}_{\ cd}\delta E^{d}_{k}=0,
\nonumber\\
&&
\epsilon^{d}_{\ cn}\delta A^{n}_{d}
=-\frac{1}{\bar{p}}\partial_{c}\varepsilon_{1}
-\frac{1}{\bar{p}}\partial_{c}(\Delta\varepsilon_{2})
\end{eqnarray}
The diffeomorphism constraint in
term of the scalar functions $(\varepsilon_{1},\varepsilon_{2})$ and
$(\kappa_{1},\kappa_{2})$ is
\begin{eqnarray} \label{EGravityDiffeomorphismCPerturbationScalarv1}
D[N^{a}] = \frac{1}{8\pi G}\int_{\Sigma}{\rm d}^{3}x \delta
N^{c}[-2\bar{p} \partial_{c}\kappa_{1}
+ \bar{q} \partial_{c} \varepsilon_{1}
+ \bar{q} \partial_{c} (\Delta \varepsilon_{2})]\,.
\end{eqnarray}
Using these relations, the Poisson bracket (\ref{HQNbarDfCaseIScalar}) is
\begin{eqnarray} \label{HQNbarDfCaseIScalarv1}
&&\{\delta H_Q[\bar{N}],D[N^{a}]\}
\nonumber\\
&&
= \frac{1}{16\pi G}\int {\rm d}^{3}x \delta N^{c}
\Bigl\{-\left(2\frac{\sigma}{\bar{q}} + 3\slashed{\sigma} + \mu + 2\nu\right)
\bar{q} \partial_{c}\varepsilon_{1}
\nonumber\\
&&
-\left(2\frac{\sigma}{\bar{q}}
+5\slashed{\sigma}
+3\mu+2\nu
+12\widetilde{\slashed{\sigma}}
+8\bar{q}\slashed{\mu}
+4\bar{q}\slashed{\nu}\right)\bar{q}\partial_{c}(\Delta \varepsilon_{2})
\nonumber\\
&&
-2\Bigl[
2\left(\frac{\partial\sigma}{\partial \bar{q}}
+3\bar{q}\frac{\partial\slashed{\sigma}}{\partial \bar{q}}
+\bar{q} \frac{\partial\mu}{\partial \bar{q}}
+2\bar{q} \frac{\partial\nu}{\partial \bar{q}}
+3\bar{q}^{2}\frac{\partial\slashed{\mu}}{\partial \bar{q}}
+\bar{q}^{2}\frac{\partial\rho}{\partial \bar{q}}
+9\bar{q}^{2}\frac{\partial \widetilde{\slashed{\sigma}}}{\partial \bar{q}}
+6\bar{q}^{2}\frac{\partial \slashed{\nu}}{\partial \bar{q}}\right)
\nonumber\\
&&
+3\slashed{\sigma}+\mu+2\nu
+6\bar{q}\slashed{\nu}
\Bigr]\bar{p}\partial_{c}\delta \kappa_{1}
+2(\slashed{\sigma}+\mu
+12\bar{q}\tilde{\slashed{\sigma}}
+6\bar{q}\slashed{\mu}
+4\bar{q}\slashed{\nu}
)\bar{p}\partial_{c}(\Delta \kappa_{2})
\Bigr\}\,.
\end{eqnarray}

This contribution would vanish classically, but may be non-zero here as long
as the scalar modes can be combined in the right form to produce a multiple of
the diffeomorphism constraint
(\ref{EGravityDiffeomorphismCPerturbationScalarv1}).
Comparing the Poisson bracket (\ref{HQNbarDfCaseIScalarv1}) with the
expression of the diffeomorphism constraint
(\ref{EGravityDiffeomorphismCPerturbationScalarv1}), we observe that
the conditions
\begin{eqnarray}\label{HDAnomalytermcancel1CaseIScalar1}
&&
\slashed{\sigma}=-\mu-2\bar{q}\slashed{\mu}\,,
\
\\
&&
\label{HDAnomalytermcancel1CaseIScalar2}
\widetilde{\slashed{\sigma}}=-\frac{1}{3}(\slashed{\mu}+\slashed{\nu})\,,
\\
&&
\label{HDAnomalytermcancel1CaseIScalar3}
\frac{\partial\sigma}{\partial \bar{q}}
+3\bar{q}\frac{\partial\slashed{\sigma}}{\partial \bar{q}}
+\bar{q} \frac{\partial\mu}{\partial \bar{q}}
+2\bar{q} \frac{\partial\nu}{\partial \bar{q}}
+3\bar{q}^{2}\frac{\partial\slashed{\mu}}{\partial \bar{q}}
+\bar{q}^{2}\frac{\partial\rho}{\partial \bar{q}}
+9\bar{q}^{2}\frac{\partial \widetilde{\slashed{\sigma}}}{\partial \bar{q}}
+6\bar{q}^{2}\frac{\partial \slashed{\nu}}{\partial \bar{q}}
\nonumber\\
&&
=-\frac{\sigma}{\bar{q}} - 6\slashed{\sigma} - 4\mu - 2\nu
-27\bar{q}\widetilde{\slashed{\sigma}}
-15\bar{q}\slashed{\mu}
-12\bar{q}\slashed{\nu}
\end{eqnarray}
imply a closed Poisson bracket:
\begin{eqnarray}
\label{HEGQNDEGNaEqDEGNA}
\{H_Q[\bar{N}],D[N^{a}]\}
= -\left(\frac{\sigma}{\bar{q}} -\mu + \nu - 3\bar{q} \slashed{\mu}
\right)D[N^{a}]\,.
\end{eqnarray}
Substituting Eqs.~(\ref{HDAnomalytermcancel1CaseIScalar1}) and
(\ref{HDAnomalytermcancel1CaseIScalar2}) into
Eq.~(\ref{HDAnomalytermcancel1CaseIScalar3}),
\begin{eqnarray}\label{HDAnomalytermcancel1CaseIScalar4}
&&
\frac{\partial \sigma}{\partial \bar{q}}
- 2 \bar{q}\frac{\partial \mu}{\partial \bar{q}}
+ 2 \bar{q}\frac{\partial \nu}{\partial \bar{q}}
-6\bar{q}^{2}\frac{\partial \slashed{\mu}}{\partial \bar{q}}
+ \bar{q}^{2} \frac{\partial \rho}{\partial \bar{q}}
+3\bar{q}^{2}\frac{\partial \slashed{\nu}}{\partial \bar{q}}
\nonumber\\
&&
=-\frac{\sigma}{\bar{q}}
+2\mu
-2 \nu
+12\bar{q}\slashed{\mu}
-3\bar{q}\slashed{\nu}\,.
\end{eqnarray}

Similarly to the case of vector modes \cite{VectorSecond}, we find that the
form of the Poisson bracket $\{H_Q[\bar{N}],D[N^{a}]\}$ may be modified by
holonomy terms.  One of the conditions for scalar modes,
Eq.~(\ref{HDAnomalytermcancel1CaseIScalar3}), is the same as that for vector
modes (Eq.~(38) in \cite{VectorSecond}).  For scalar modes, however, we need
the additional conditions (\ref{HDAnomalytermcancel1CaseIScalar1}) and
(\ref{HDAnomalytermcancel1CaseIScalar2}).  The requirement of
having anomaly-free constraints therefore imposes tighter restrictions on the
parameters of the holonomy-modification function (\ref{ficdconcrete}) when we
consider scalar modes.

Moreover, for the contribution (\ref{HQdeltaNDScalar}) to the Poisson bracket,
we have
\begin{eqnarray} \label{HQdeltaNDfCaseIScalar}
&&\{\delta H_Q[\delta N],D[N^{a}]\}
\nonumber\\
&&
=\delta H_Q[\delta N^{c}\partial_{c}\delta N]
-\frac{1}{16\pi G}\int d^{3}x (\delta N^{c}\partial_{c}\delta N) 4 \bar{q}
\sqrt{\bar{p}}\left(\frac{\sigma}{\bar{q}}-\mu+\nu-
3\slashed{\mu}\bar{q}\right)
\end{eqnarray}
using Eqs.~(\ref{HDAnomalytermcancel1CaseIScalar1}) and
(\ref{HDAnomalytermcancel1CaseIScalar2}).  The condition of anomaly-free
constraints requires
\begin{eqnarray} \label{HDAnomalytermcancel2CaseIScalar}
\frac{\sigma}{\bar{q}}-\mu+\nu-3\slashed{\mu}\bar{q}=0\,,
\end{eqnarray}
so that $\{\delta H_Q[N],D[N^{a}]\}$ in (\ref{HEGQNDEGNaEqDEGNA})
vanishes.  Therefore, when the conditions
(\ref{HDAnomalytermcancel1CaseIScalar1}),
(\ref{HDAnomalytermcancel1CaseIScalar2}),
(\ref{HDAnomalytermcancel1CaseIScalar4}) and
(\ref{HDAnomalytermcancel2CaseIScalar}) are satisfied, the Poisson bracket
between the holonomy-modified Hamiltonian and diffeomorphism constraints
becomes
\begin{eqnarray}
\{\delta H_Q[N],D[N^{a}]\}
=\delta H_Q[\delta N^{c}\partial_{c}\delta N],
\end{eqnarray}
which is identical to the classical case.  The conditions
(\ref{HDAnomalytermcancel1CaseIScalar1}),
(\ref{HDAnomalytermcancel1CaseIScalar2}),
(\ref{HDAnomalytermcancel1CaseIScalar4}) and
(\ref{HDAnomalytermcancel2CaseIScalar}) can be combined as
\begin{eqnarray} \label{HDAnomalytermcancel1TotalScalar}
&&
\sigma=\bar{q}\mu-\bar{q}\nu+3\bar{q}^{2}\slashed{\mu}\,,
\nonumber\\
&&
\slashed{\sigma}=-\mu-2\bar{q}\slashed{\mu}\,,
\nonumber\\
&&
\widetilde{\slashed{\sigma}}=-\frac{1}{3}(\slashed{\mu}+\slashed{\nu})\,,
\nonumber\\
&&
- \bar{q}\frac{\partial \mu}{\partial \bar{q}}
+ \bar{q}\frac{\partial \nu}{\partial \bar{q}}
-3\bar{q}^{2}\frac{\partial \slashed{\mu}}{\partial \bar{q}}
+ \bar{q}^{2} \frac{\partial \rho}{\partial \bar{q}}
+3\bar{q}^{2}\frac{\partial \slashed{\nu}}{\partial \bar{q}}
=3\bar{q}\slashed{\mu}
-3\bar{q}\slashed{\nu}\,.
\end{eqnarray}

\subsection{Two Hamiltonian constraints}

We now turn to the study of the Poisson bracket between two holonomy-modified
Hamiltonian constraints.  Using Eq.~(\ref{HDAnomalytermcancel1TotalScalar}),
the terms of the holonomy modification function (\ref{ficdexpandCaseIScalar})
can be rewritten as
\begin{eqnarray} \label{ficdexpandCaseIbeforeHHScalar}
f^{i(0)}_{cd} &=& (- \mu \bar{q} +\nu \bar{q}
-3\slashed{\mu}\bar{q}^{2}+3\slashed{\nu}\bar{q}^{2}
+ \rho\bar{q}^{2})\epsilon^{i}_{\ cd}\,,
\nonumber\\
\mathcal{A}^{i(1)}_{cd} &=&
(\nu+\rho\bar{q}+3\slashed{\nu}\bar{q})
(\epsilon^{i}_{\ md}\delta A^{m}_{c}+\epsilon^{i}_{\ cn}\delta A^{n}_{d})
-(\mu+3\slashed{\mu}\bar{q})\delta A^{j}_{a}\delta^{a}_{j}\epsilon^{i}_{\ cd}
+(\mu+3\slashed{\mu}\bar{q})\delta A^{i}_{b}\epsilon^{b}_{\ cd}\,,
\nonumber\\
\mathcal{B}^{i(1)}_{cd} &=& \tau \partial_{[c}\delta A^{i}_{d]}\,,
\nonumber\\
\mathcal{A}^{i(2)}_{cd} &=&
\rho\epsilon^{i}_{\ mn}\delta A^{m}_{c} \delta A^{n}_{d}
-\frac{1}{3}(\slashed{\mu}+\slashed{\nu})(\delta
A^{j}_{a}\delta^{a}_{j})^{2}\epsilon^{i}_{\ cd}
+\slashed{\nu}(\epsilon^{i}_{\ md}\delta A^{m}_{c}+\epsilon^{i}_{\ cn}\delta
A^{n}_{d})\delta A^{j}_{a}\delta^{a}_{j}
+\slashed{\mu} \delta A^{i}_{b}\epsilon^{b}_{\ cd}\delta A^{j}_{a}\delta^{a}_{j}\,,
\nonumber\\
\mathcal{B}^{i(2)}_{cd}&=&0\,,
\nonumber\\
\mathcal{C}^{i(2)}_{cd}&=&
0\,.
\end{eqnarray}
In terms of
\begin{eqnarray} \label{ficdexpandNecessaryRelationCaseIbeforeHHScalar}
&&
\mathcal{F}^{(0)} = 6(- \mu \bar{q} +\nu \bar{q}
-3\slashed{\mu}\bar{q}^{2}+3\slashed{\nu}\bar{q}^{2}
+ \rho\bar{q}^{2})\,,
\nonumber\\
&&
\mathcal{A}^{(1)} =
(-4\mu+4\nu
-12\slashed{\mu}\bar{q}+12\slashed{\nu}\bar{q}
+4\rho\bar{q}
)\delta A^{k}_{d}\delta^{d}_{k}
=\frac{2}{3}\frac{\mathcal{F}^{(0)}}{\bar{q}}\delta A^{k}_{d}\delta^{d}_{k}\,,
\nonumber\\
&&
\mathcal{A}^{(2)} =
\rho(\delta A^{k}_{d}\delta^{d}_{k})^{2}
-\rho \delta A^{m}_{c}\delta A^{n}_{d}\delta^{c}_{n}\delta^{d}_{m} \,,
\nonumber\\
&&
\frac{\partial \mathcal{F}^{(0)}}{\partial \bar{q}} =
6(-\mu+\nu-3\slashed{\mu}\bar{q}+3\slashed{\nu}\bar{q}+2\rho\bar{q})
=\frac{\mathcal{F}^{(0)}}{\bar{q}}
+6\rho \bar{q}\,,
\nonumber\\
&&
f^{i(0)}_{cd}=\frac{\mathcal{F}^{(0)}}{6}\epsilon^{i}_{\ cd}\quad,\quad
\mathcal{B}^{(1)} = \tau \partial_{c}\delta A^{i}_{d}\epsilon^{\ cd}_{i}\,,
\nonumber\\
&&
\frac{\partial \mathcal{A}^{(1)}}{\partial \bar{q}}=4\rho \delta
A^{k}_{d}\delta^{d}_{k} \,,
\end{eqnarray}
the sum of the holonomy-modified Poisson brackets, (\ref{PBCQ}), (\ref{PBQC})
and (\ref{PBtwoHCQScalar}) is
\begin{eqnarray}
\label{PBtwoHCQCaseIScalar}
&&
\{H[\delta N_{1}-\delta N_{2}],\delta H_Q[\bar{N}]\}
+\{\delta H_Q[\delta N_{1}-\delta N_{2}],H[\bar{N}]\}+\{\delta H_Q[\delta
N_{1}-\delta N_{2}],\delta H_Q[\bar{N}]\}
\nonumber\\
&=&
\frac{1}{8\pi G}\int {\rm d}^{3}x \partial^{c}(\delta N_{2}-\delta
N_{1})\frac{\bar{N}}{\bar{p}}
\Bigl\{-(\frac{1}{4}\tau^{2}+\tau)
\left[\bar{p}\partial_{k}\delta A^{k}_{c}-\bar{p}\partial_{c}(\delta
  A^{k}_{d}\delta^{d}_{k})\right]
\nonumber\\
&&
+\left[-\frac{1}{24}(\mathcal{F}^{(0)}+6\bar{q}^2)\frac{\partial\tau}{\partial \bar{q}}
+(1+\frac{\tau}{2})(\mu+3\slashed{\mu}\bar{q})
+\frac{1}{6}\frac{\mathcal{F}^{(0)}}{\bar{q}}
+\frac{1}{2}\tau \bar{q}
+\frac{1}{4}\tau^2 \bar{q}
\right]\bar{p}\delta A^{n}_{d}\epsilon^{d}_{\ cn}
\nonumber\\
&&
-\frac{1}{12}(\tau\mathcal{F}^{(0)}+2\mathcal{F}^{(0)}+6\tau\bar{q}^2)
\epsilon^{k}_{\ cd}\delta E^{d}_{k}
\Bigr\}
\nonumber\\
&&
+\frac{1}{8\pi G}\int {\rm d}^{3}x (\delta N_{2}-\delta N_{1})
\frac{1}{72}(\mathcal{F}^{(0)}+6\bar{q}^{2})(\mathcal{F}^{(0)}-6\rho \bar{q}^{2})
\left(\frac{1}{\bar{q}\sqrt{\bar{p}}}\delta E^{d}_{k}\delta^{k}_{d}
-2\frac{\sqrt{\bar{p}}}{\bar{q}^{2}}\delta A^{k}_{d}\delta^{d}_{k}
\right)\,.
\end{eqnarray}
By using Eqs.~(\ref{SomeRelationScalarConcrete}), we express the
Poisson bracket (\ref{PBtwoHCQCaseIScalar}) in terms of the scalar modes
$(\varepsilon_{1},\varepsilon_{2})$ and $(\kappa_{1},\kappa_{2})$ as
\begin{eqnarray}
\label{PBtwoHCQCaseIScalarv1}
&&\{H[\delta N_{1}-\delta N_{2}],\delta H_Q[\bar{N}]\}
+\{\delta H_Q[\delta N_{1}-\delta N_{2}],H[\bar{N}]\}+\{\delta H_Q[\delta
N_{1}-\delta N_{2}],\delta H_Q[\bar{N}]\}
\nonumber\\
&=&
\frac{1}{8\pi G}\int {\rm d}^{3}x \partial^{c}(\delta N_{2}-\delta N_{1})\frac{\bar{N}}{\bar{p}}
\Bigl\{(\frac{1}{2}\tau^{2}+2\tau)\bar{p}\partial_{c}\kappa_{1}
\nonumber\\
&&
-\left[-\frac{1}{24}(\mathcal{F}^{(0)}+6\bar{q}^2)\frac{\partial\tau}{\partial \bar{q}}
+(1+\frac{\tau}{2})(\mu+3\slashed{\mu}\bar{q})
+\frac{1}{6}\frac{\mathcal{F}^{(0)}}{\bar{q}}
+\frac{1}{2}\tau \bar{q}
+\frac{1}{4}\tau^2 \bar{q}
\right]
\partial_{c}(\varepsilon_{1}+\Delta\varepsilon_{2})
\Bigr\}
\nonumber\\
&&
+\frac{1}{8\pi G}\int {\rm d}^{3}x (\delta N_{2}-\delta N_{1})
\frac{1}{72}(\mathcal{F}^{(0)}+6\bar{q}^{2})(\mathcal{F}^{(0)}-6\rho \bar{q}^{2})
\left[\frac{1}{\bar{q}\sqrt{\bar{p}}}(3\varepsilon_{1}+\Delta \varepsilon_{2})
-2\frac{\sqrt{\bar{p}}}{\bar{q}^{2}}(3\kappa_{1}+\Delta \kappa_{2})
\right]\,.
\nonumber\\
\end{eqnarray}
Equation (\ref{PBtwoHCQCaseIScalarv1}) implies that, in order to have a closed
Poisson bracket, we should impose the conditions
\begin{eqnarray}
\label{PBtwoHCQCancellCaseIScalarv3}
&&
(\rho+1) \bar{q}\frac{\partial \tau}{\partial \bar{q}}
-4\left(1+\frac{\tau}{2}\right)\left(\frac{\mu}{\bar{q}}+3\slashed{\mu}\right)
+2\tau-4\rho
=0,
\nonumber\\
&&
\mu=\nu-3\slashed{\mu}\bar{q}+3\slashed{\nu}\bar{q}\,.
\end{eqnarray}
The Poisson bracket (\ref{PBtwoHCQCaseIScalarv1}) can then be expressed as
\begin{eqnarray} \label{DefAlg}
&&\{H[\delta N_{1}-\delta N_{2}],\delta H_Q[\bar{N}]\}
+\{\delta H_Q[\delta N_{1}-\delta N_{2}],H[\bar{N}]\}+\{\delta H_Q[\delta
N_{1}-\delta N_{2}],\delta H_Q[\bar{N}]\}
\nonumber\\
&=&-\left(\frac{1}{4}\tau^{2}+\tau\right)
D[\bar{N}\bar{p}^{-1}\partial^{c}(\delta
N_{2}-\delta N_{1})]\,.
\end{eqnarray}
Using (\ref{PBtwoHCQCancellCaseIScalarv3}), we obtain
\begin{equation}
 -\bar{q}\frac{\partial\mu}{\partial\bar{q}}+
 \bar{q}\frac{\partial\nu}{\partial\bar{q}}= \bar{q}\frac{\partial}{\partial
   \bar{q}}\left(3\bar{q}(\slashed{\mu}-\slashed{\nu})\right)= 3\bar{q}^2
   \frac{\partial(\slashed{\mu}-\slashed{\nu})}{\partial\bar{q}}+
   3\bar{q}(\slashed{\mu}-\slashed{\nu})
\end{equation}
such that (\ref{HDAnomalytermcancel1TotalScalar}) simplifies to
$\bar{q}^2\partial\rho/\partial\bar{q}=0$. We arrive at the
conditions
\begin{eqnarray} \label{TotalConditionCaseIScalarv3}
&&
\rho=c_1\,,
\nonumber
\\
&&
(c_1+1) \bar{q}\frac{\partial \tau}{\partial \bar{q}}
-4\left(1+\frac{\tau}{2}\right)\left(\frac{\mu}{\bar{q}}+3\slashed{\mu}\right)
+2\tau-4 c_1
=0\,,
\nonumber\\
&&
\sigma=3\bar{q}^{2}\slashed{\nu}\,,
\nonumber\\
&&
\slashed{\sigma}=-\nu+ \bar{q}\slashed{\mu}-3\bar{q}\slashed{\nu}\,,
\nonumber\\
&&
\widetilde{\slashed{\sigma}}=-\frac{1}{3}(\slashed{\mu}+\slashed{\nu})\,,
\nonumber\\
&&
\mu=\nu-3\slashed{\mu}\bar{q}+3\slashed{\nu}\bar{q}\,,
\end{eqnarray}
on anomaly-free constraints, where $\rho=c_1$ is now a constant
independent of $\bar{q}$.  In these conditions, there are three free
functions of $\bar{q}$: $\nu$, $\slashed{\mu}$ and $\slashed{\nu}$.

In \cite{VectorSecond}, a different-looking equation, (38), has been derived
for anomaly-freedom of vector modes. Slightly adapted to our notation, this
condition reads
\begin{eqnarray}
 0&=&\frac{\partial\sigma}{\partial\bar{q}}+
 3\bar{q}\frac{\partial\slashed{\sigma}}{\partial\bar{q}}+
 9\bar{q}^2\frac{\partial\widetilde{\slashed{\sigma}}}{\partial\bar{q}}+
 \bar{q}\frac{\partial\mu}{\partial\bar{q}}+
 2\bar{q}\frac{\partial\nu}{\partial\bar{q}}+
 3\bar{q}^2\frac{\partial\slashed{\mu}}{\partial\bar{q}}+
 6\bar{q}^2\frac{\partial\slashed{\nu}}{\partial\bar{q}}+
 \bar{q}^2\frac{\partial\rho}{\partial\bar{q}}\nonumber\\
&& +\frac{\sigma}{\bar{q}}+
6\slashed{\sigma}+27\bar{q}\widetilde{\slashed{\sigma}}+
4\mu+15\bar{q}\slashed{\mu}+ 2\nu+ 12\bar{q}\slashed{\nu}\,.
\end{eqnarray}
If we insert (\ref{TotalConditionCaseIScalarv3}), this equation is identically
satisfied, such that the formulations for scalar and vector modes are
consistent with each other.

We have found a candidate for a non-trivial holonomy-modified function
$f^{i}_{cd}$, which satisfies anomaly-free constraint brackets for both scalar
and vector modes up to second order.  This non-trivial function can be
written as
\begin{eqnarray}\label{ficdconcreteFinalSecondOrderScalar}
f^{i}_{cd}&=&3\bar{q}^{2}\slashed{\nu}\epsilon^{i}_{\ cd}\nonumber\\
&&+(-\nu+ \bar{q}\slashed{\mu}-3\bar{q}\slashed{\nu})\epsilon^{i}_{\ cd}
A^{j}_{a}\delta^{a}_{j}
+(\nu-3\slashed{\mu}\bar{q}+3\bar{q} \slashed{\nu}) A^{i}_{b}
\epsilon^{b}_{\ cd}
+\nu(\epsilon^{i}_{\ md} A^{m}_{c} + \epsilon^{i}_{\ cn} A^{n}_{d})
\nonumber\\
&&-\frac{1}{3}(\slashed{\mu}+ \slashed{\nu})\epsilon^{i}_{\ cd}
(A^{j}_{a}\delta^{a}_{j})^{2}
+\slashed{\mu} A^{i}_{b}\epsilon^{b}_{\ cd} A^{j}_{a}\delta^{a}_{j}
+\slashed{\nu}(\epsilon^{i}_{\ md} A^{m}_{c} +
\epsilon^{i}_{\ cn} A^{n}_{d}) A^{j}_{a}\delta^{a}_{j}
+ c_1\epsilon^{i}_{\ mn} A^{m}_{c} A^{n}_{d}\nonumber\\
&&+ \tau\partial_{[c} A^{i}_{d]} \,.
\end{eqnarray}
Here, $\tau$ is determined by the second equation in
Eq.~(\ref{TotalConditionCaseIScalarv3}). When
$\nu=\slashed{\mu}=\slashed{\nu}=0$, $\tau=2c_1$, the modification function
returns to the form of classical curvature as $f^{i}_{cd}=c_1 F^{i}_{cd}$, in which
$\rho=c_1$ is a constant and can be absorbed in the definition of $G$.

\subsection{SU(2)-covariance}

It remains to check the SU(2)-covariance of the holonomy-modification function
$f^{i}_{cd}$ in (\ref{ficdconcreteFinalSecondOrderScalar}).  To this end, we
calculate the Poisson bracket between the holonomy modifications of the
Hamiltonian constraint, $\delta H_Q[N]$, and the Gauss constraint
$G[\Lambda]$:
\begin{align}
&
\{\delta H_Q[N],G[\Lambda]\}
\nonumber
\\
&=
\frac{1}{16\pi G}\int d^3x N\sqrt{|\det E|}
\Big[4\bar{q}\slashed{\nu}(3-A_b^k\delta^b_k)D_l\Lambda^l
+
(\tau-2c_1)
(A_b^ke^b_kD_l\Lambda^l -A^k_be^b_lD_k\Lambda^l)
\nonumber
\\
&
+
2\Lambda^l
\Big(
(2\nu-3\slashed{\mu}\bar{q}+3\bar{q} \slashed{\nu})A^k_b\epsilon^b_{\ lk}
+(\slashed{\mu}+\slashed{\nu})A^j_a\delta^a_jA^k_b\epsilon^b_{\ lk}
+\Big(c_1-\frac{1}{2}\tau\Big)\epsilon^d_{\ mn}A^m_cA^n_de^c_l
\Big)\Big]
\,.
\label{HGp}
\end{align}
We have introduced the covariant derivative defined as
\begin{equation}
D_av^i=\partial_av^i-\epsilon^i_{\ jk}A^j_av^k\,.
\end{equation}

From Eq.~(\ref{HGp}), it is easy to conclude that the Poison bracket $\{\delta
H_Q[N],G[\Lambda]\}$ vanishes only if the parameters satisfy
$\nu=\slashed{\mu}=\slashed{\nu}=0$, $\tau=2c_1$.  The modification function
then returns to the classical case of $f^{i}_{cd}\propto F^{i}_{cd}$.
Therefore, if we now combine the constraint brackets with the condition that
all expressions be invariant under SU(2) transformations, the system turns out
to be strongly restricted: In (\ref{ficdconcreteFinalSecondOrderScalar}), only
the last two terms (with coefficients $\rho$ and $\tau$) can appear in an
SU(2)-covariant expression, as is well known from the possible covariant
combinations of connection components. Moreover, the combination of the last
two terms is covariant only if $\tau=2\rho=2c_1$. All other terms in
(\ref{ficdconcreteFinalSecondOrderScalar}) which are quadratic in $\delta
A_a^i$ must be zero, so that $\slashed{\mu}=\slashed{\nu}=\nu=0$, and also the
first background contribution is ruled out. In particular, background holonomy
modifications are ruled out in this model, which would give rise to a function
$\slashed{\nu}(\bar{q})=(\ell\bar{q})^{-2}\sin^2(\ell\bar{q})-1\not=0$. This
result is in contrast to \cite{ScalarHolInv}, where a consistent version with
background holonomy modifications has been found using an extrinsic-curvature
formulation instead of a connection formulation. (If there were background
holonomy modifications similar to \cite{ScalarHolInv}, they should contribute
to (\ref{DefAlg}) a factor of $\partial^2\slashed{\nu}/\partial\bar{q}^2$ in
addition to $\tau^2$.)  The only allowed correction here is a function $\tau$
which would multiply the classical $F_{cd}^i$. Such a modification resembles
the results from inverse-triad rather than holonomy modifications.

The appearance of this modification, however, shows an interesting analogy
with the results of \cite{ScalarHolInv}: A crucial factor in the deformation
function found in this paper is called $1+\tau_3$ there, which is the
coefficient of $\partial_c\partial^j\delta E_j^c$ in the linear term of the
Hamiltonian constraint. Such a second derivative of the triad perturbation
also appears here, when one writes $\partial_{[c}\delta A_{d]}^i$ in terms of
the spin connection and extrinsic curvature, and this term in
(\ref{ficdconcreteFinalSecondOrderScalar}) has the coefficient that appears in
the deformation function in (\ref{DefAlg}).  Also the deformed brackets
(\ref{DefAlg}) resemble those found for inverse-triad corrections: The
deformation function in the full contraint $H+\delta H_Q$ is one plus a
correction function which does not necessarily change sign. There is an
indication that signature change may be avoided in the presence of holonomy
modifications because the deformation function in (\ref{DefAlg}) does not
depend on $\slashed{\nu}$, but then holonomy modifications are ruled out
altogether in this model.

\section{Conclusions}\label{Conclusions}

We have extended the investigation of inhomogeneous perturbations of the
effective Hamiltonian constraint with holonomy modifications in Euclidean
models of loop quantum gravity to include scalar modes.  The Poisson brackets
between a holonomy-modified Hamiltonian constraint and the diffeomorphism
constraint as well as that between the two holonomy-modified Hamiltonian
constraints have been calculated. It turns out that anomaly-free scalar modes
impose stronger restrictions on the parameters of the holonomy-modification
function than vector modes, but non-trivial modifications remain possible such
that the Poisson brackets of Hamiltonian and diffeomorphism constraints are
anomaly-free.  If SU(2)-covariance is implemented, however, the modifications
are much more tightly restricted, even ruling out background holonomy
modifications. These results have several new implications and help to clarify
relationships between previous studies.

\subsection{Spatial derivatives}

The main new ingredient used here, compared with existing models which allow
background holonomy modifications, is the possibility of new corrections even
at the classical form of at most first-order derivatives of the connection. We
have motivated these new terms by starting with a connection rather than
extrinsic-curvature formulation, in which case the derivative structure of the
Hamiltonian constraint is different. The appearance of derivatives, in turn,
affects possible modifications of constraint brackets derived using
integration by parts.

If one does not implement SU(2)-covariance, one does not obtain a physical
model but may still consider algebraic aspects of the system of Hamiltonian
and diffeomorphism constraints, which turns out to be quite non-trivial. In
this case, there are several free coefficients in
(\ref{ficdconcreteFinalSecondOrderScalar}), including background holonomy
modifications $\slashed{\nu}$. However, the resulting bracket (\ref{DefAlg})
does not show the characteristic form found in other such cases, which have
led to signature change at high density: Structure functions in the classical
bracket would be multiplied by
$\frac{1}{2}\partial^2\slashed{\nu}/\partial\bar{q}^2$ for this form to be
realized, but we have seen no such factor.

Although our results do not provide a physical model in this case, they may
indicate that it is possible to avoid signature change and the associated
indeterministic behavior, provided one starts with a connection
formulation. The appearance of spatial derivatives in the Hamiltonian
constraint is then different from an extrinsic-curvature formulation, which
can affect the constraint brackets when integrating by parts. An
extrinsic-curvature formulation has a classical Hamiltonian constraint without
derivatives of the extrinsic curvature, while the densitized triad appears
with up to second-order derivatives. In a connection formulation, the
connection appears with up to first-order derivatives, while the densitized
triad does not have derivative terms in the version used here, that is with
$\gamma=1$. The fact that signature change appears in the former but not in
the latter case is consistent with the simple 1-dimensional model of
\cite{Loss}.

In a comparison with results from self-dual variables
\cite{SphSymmComplex,CosmoComplex,GowdyComplex}, we see similar properties in
that signature change or, more generally, modifications of the constraint
brackets do not seem generic. Also at a formal level there are similarities,
in particular the appearance of spatial derivatives of the connection in the
constraint, which do not appear in extrinsic-curvature versions, and a more
important role played by the Gauss constraint. The latter is usually solved
explicitly in extrinsic-curvature formulations, which automatically ensures
compatibility with its flow but also leads to less ambiguity in identifying
the Hamiltonian constraint. In a connection formulation, by contrast, the
Hamiltonian constraint is defined only up to multiples of the Gauss
constraint. The form (\ref{GravityHamiltonian}) used here is conventional, but
not unique. One could use the Gauss constraint in order to eliminate spatial
deirvatives of the connection, which may bring the structure closer to an
extrinsic-curvature formulation. Such brackets, however, are beyond the scope
of the present paper.

Since SU(2)-covariance leads to significant restrictions of the allowed
modifications, the form of holonomies as covariant functionals of the
connection gives further indications as to how a fully anomaly-free system
could be found in a connection formulation. Holonomies are non-local in space
because they are computed by integrating the connection over a curve. In an
effective theory, such an expression appears in the form of a derivative
expansion that does not end at any finite order. Therefore, SU(2)-covariant
formulations may require higher spatial derivatives beyond the classical
order. Holonomies as used in kinematical constructions of loop quantum gravity
suggest that higher spatial derivatives are unaccompanied by higher time
derivatives because one uses only spatial curves in holonomies. However, this
picture suggests problems with space-time covariance because it is difficult
to maintain different orders of space and time derivatives in a covariant
formulation or, alternatively, because a spatial curve embedded in space-time
may no longer be spatial after a general coordinate transformation.

\subsection{Comparison between effective and operator approaches}

The preceding arguments provide intuitive reasons why it seems difficult to
have space-time covariance and SU(2) covariance in the same holonomy-based
theory. At a formal level, these difficulties have been confirmed in
spherically symmetric models \cite{HigherSpatial}. On the other hand,
\cite{OffShell,ConstraintsG} suggest that very careful routings of loops used
to construct holonomies for a quantization of the Hamiltonian constraint could
lead to a full anomaly-free quantum theory. A comparison between these results
therefore seems useful.

The operator constructions of \cite{OffShell,ConstraintsG} so far have not
given indications about possible deformations of the constraint
brackets. Since they have been obtained in Euclidean gravity, our present
results help to reconcile this outcome with those of effective derivations
based on real variables, which generically lead to deformed constraint
brackets. In the present paper, we used Euclidean gravity with a derivative
structure of the Hamiltonian constraint that is more similar to the
constraints quantized in \cite{OffShell,ConstraintsG} than those of effective
approaches in real variables. And here, as in the case of self-dual
connections \cite{SphSymmComplex,CosmoComplex,GowdyComplex}, the constraint
brackets are subject to different modifications compared with Lorentzian
models in real variables, and no deformations are possible if SU(2)-covariance
is imposed.

While these are qualitative similarities, we emphasize that a comparison of
constraint brackets in effective and operator approaches is not
straightforward. Effective approaches, by construction, lead to constraints
and brackets of classical type, and therefore implicitly assume that there is
an underlying semiclassical state in which one has taken expectation
values. Using the systematic treatment of canonical effective constraints
\cite{EffAc,Karpacz,EffCons,EffConsRel,EffConsQBR}, one can derive properties
of such a semiclassical state within the effective formalism, but not much has
been done in this direction in spherically symmetric or perturbed cosmological
models. The operator treatment, on the other hand, results in commutators
instead of brackets. A detailed comparison would therefore require an
understanding of the semiclassical limit of loop quantum gravity, perhaps with
input from effective results about semiclassical states. Consistent versions
of effective constraint brackets should then be compared with expectation
values of consistent commutators of constraint operators computed in a
semiclassical state. Only the latter step of computing semiclassical
expectation values would give unambiguous results about possible deformations
of constraint brackets in operator approaches. Unfortunately, no such results
are available owing to the complicated nature of the semiclassical limit of
loop quantum gravity.

\subsection{Space-time structure}

The question whether it is possible to avoid indeterministic behavior in
effective models of loop quantum gravity remains open, but at least the
present results have confirmed the indications of
\cite{SphSymmComplex,CosmoComplex,GowdyComplex} pointing to an affirmative
answer. The form of signature change appears to depend on the specific
formulation used, so that its absence would provide an additional restrictive
condition together with anomaly-freedom alone. However, existing results need
to be extended in several directions before a firm conclusion can be
drawn. First, the derivative nature of the Hamiltonian constraint and
therefore the constraint brackets are different for $\gamma\not=1$, even if
one still considers Euclidean gravity.  Second, spatial derivatives in an
effective constraint may be generated by quantum corrections even if they are
absent from the classical constraint. All such derivatives should be included
unless they are prohibited by symmetries. In background-independent quantum
theories of gravity, symmetries of space-time are to be derived and do not
restrict the terms in effective constraints used before anomaly-free brackets
have been obtained. For the same reason, our calculations should be extended
by including a general derivative expansion not just of $A_a^i$ but also of
$E^b_j$.  In the same vein, one should extend our setup in this paper to the
Lorentzian case, where the construction of the effective holonomy-modified
Hamiltonian of full loop quantum gravity and the calculation of constraint
brackets would be more complicated, as even the classical constraint would
contain spatial derivatives of $E^b_j$ via $\Gamma_a^i$. Matter terms added in
an anomaly-free way provide another large question, as do the possible forms
of higher-derivative corrections in both space and time.

There are therefore several extensions of existing calculations which should
be completed before reliable conclusions about the potential consistency of
loop quantum gravity can be drawn. Our present results differ in some crucial
respects from previous calculations and should therefore help to provide a
better estimate of the options realized in models of cosmological
perturbations within this framework.

\begin{acknowledgments}

  This work is supported by the Natural Science Foundation of China
  (NSFC) (Grant Nos.~11775036, 11475023 and 11875006), NSF grant
  PHY-1607414, J.~P.\ Wu is also supported by Natural Science
  Foundation of Liaoning Province under Grant No.~201602013.

\end{acknowledgments}


\end{document}